\documentclass[twocolumn,superscriptaddress]{revtex4-2}

\usepackage{graphicx}
\usepackage{dcolumn}
\usepackage{bm}

\usepackage{color}
\usepackage{amsmath}
\usepackage{float}
\usepackage{booktabs}
\usepackage{caption}
\usepackage{ragged2e}  
\DeclareUnicodeCharacter{2550}{\ensuremath{\equiv}}
\usepackage[pdftex]{hyperref}
\definecolor{darkred}{rgb}{0.25,0,0}
\definecolor{darkgreen}{rgb}{0,0.25,0}
\definecolor{darkblue}{rgb}{0,0,1}
\hypersetup{colorlinks,linkcolor=darkblue,filecolor=black,urlcolor=darkblue,citecolor=darkblue}
\usepackage{multirow}
\usepackage{textcomp}
\usepackage[compact]{titlesec}
\vspace{-0.5cm} 
\begin{document}

\preprint{APS/123-QED}

\title{Weak effects of electron-phonon interactions on the lattice thermal conductivity of wurtzite GaN with high electron concentrations}


\author{Jianshi Sun}
\affiliation{Institute of Micro/Nano Electromechanical System and Integrated Circuit, College of Mechanical Engineering, Donghua University, Shanghai 201620, China
}%

\author{Shouhang Li}
\email{shouhang.li@dhu.edu.cn}
\affiliation{Institute of Micro/Nano Electromechanical System and Integrated Circuit, College of Mechanical Engineering, Donghua University, Shanghai 201620, China
}%

\author{Zhen Tong}
\affiliation{
School of Advanced Energy, Sun Yat-Sen University, Shenzhen 518107, China
}

\author{Cheng Shao}
\affiliation{
 Thermal Science Research Center, Shandong Institute of Advanced Technology, Jinan, Shandong 250103, China
}

\author{Xiangchuan Chen}
\affiliation{%
Institute of Micro/Nano Electromechanical System and Integrated Circuit, College of Mechanical Engineering, Donghua University, Shanghai 201620, China
}

\author{Qianqian Liu}
\affiliation{%
Institute of Micro/Nano Electromechanical System and Integrated Circuit, College of Mechanical Engineering, Donghua University, Shanghai 201620, China
}

\author{Yucheng Xiong}
\affiliation{%
Institute of Micro/Nano Electromechanical System and Integrated Circuit, College of Mechanical Engineering, Donghua University, Shanghai 201620, China
}

\author{Meng An}
\affiliation{
 Department of Mechanical Engineering, The University of Tokyo, 7-3-1 Hongo, Bunkyo, Tokyo, 113-8656, Japan
}

\author{Xiangjun Liu}
\email{xjliu@dhu.edu.cn}
\affiliation{%
Institute of Micro/Nano Electromechanical System and Integrated Circuit, College of Mechanical Engineering, Donghua University, Shanghai 201620, China
}%

\date{\today}

\begin{abstract}
Wurtzite gallium nitride (GaN) has great potential for high-frequency and high-power applications due to its excellent electrical and thermal transport properties. However, enhancing the performance of GaN-based power electronics relies on heavy doping. Previous studies showed that electron-phonon interactions have strong effects on the lattice thermal conductivity of GaN due to the Fröhlich interaction. Surprisingly, our investigation reveals weak effects of electron-phonon interactions on the lattice thermal conductivity of \textit{n}-type GaN at ultra-high electron concentrations and the impact of the Fröhlich interaction can be ignored. The small phonon-electron scattering rate is attributed to the limited scattering channels, quantified by the Fermi surface nesting function. In contrast, there is a significant reduction in the lattice thermal conductivity of \textit{p}-type GaN at high hole concentrations due to the relatively larger Fermi surface nesting function. Meanwhile, as \textit{p}-type GaN has relatively smaller electron-phonon matrix elements, the reduction in lattice thermal conductivity is still weaker than that observed in \textit{p}-type silicon. Our work provides a deep understanding of thermal transport in doped GaN and the conclusions can be further extended to other wide-band-gap semiconductors, including $\beta$-Ga$_2$O$_3$, AlN, and ZnO.
\end{abstract}

\maketitle

\section{INTRODUCTION}
GaN has a high Baliga figure of merit and large lattice thermal conductivity, which makes it promising for power electronics\cite{roccaforte2018emerging} and optoelectronics\cite{chung2010transferable}. Generally, high charge carrier concentrations are required to obtain large electrical conductivity in GaN. For instance, when the hole concentration is increased to $2 \times 10^{18} \, \text{cm}^{-3}$ doping with magnesium and oxygen atoms, the resistivity is reduced to 0.2 $\Omega\cdot\text{cm}$\cite{korotkov2001electrical}. Similarly, for \textit{n}-type GaN, a high electrical conductivity exceeding $4 \times 10^3$ $\Omega^{-1}\cdot\text{cm}^{-1}$ is achieved with an electron concentration of $3.7 \times 10^{20} \, \text{cm}^{-3}$ through germanium doping\cite{konczewicz2022electrical}. However, high-concentration doping can have a negative influence on the lattice thermal transport. An extra phonon scattering term, namely phonon-electron scattering, is introduced when there are plenty of charge carriers. This may lead to a pronounced reduction in lattice thermal conductivity, which is harmful to the performance of power semiconductor devices\cite{warzoha2021applications}.

Thanks to the giant advancement in computational capability, it is possible to rigorously consider the electron-phonon interactions on the lattice thermal conductivity from first-principles calculations\cite{tong2019comprehensive,jain2016thermal,chen2019understanding}. Liao \emph{et al}. found that there is a significant reduction of $\sim37\%$ in lattice thermal conductivity for \textit{n}-type silicon (Si) with an electron concentration of $10^{21} \, \text{cm}^{-3}$\cite{liao2015significant}. Their predictions were later verified by the three-pulse femtosecond photoacoustic technique\cite{liao2016photo}. Similar trends were also observed in two-dimensional semiconductor materials, such as $\text{MoS}{_2}$ and PtSSe\cite{liu2020strong} and phosphorene and silicene\cite{yue2019controlling}. Notably, GaN holds a much larger lattice thermal conductivity than Si because there is a large acoustic-optical phonon frequency gap which greatly limits available phonon-phonon scattering channels\cite{lindsay2012thermal}. Given the weak phonon-phonon scattering, the phonon-electron scattering is expected to play a crucial role in GaN at high carrier concentrations, as observed in another wide-band-gap semiconductor 3C-SiC\cite{wang2017strong}. Previous studies by Yang \emph{et al}. and Tang \emph{et al}. reported that the lattice thermal conductivity of GaN is severely limited by the strong Fröhlich interaction between electrons and long-wavelength longitudinal optical phonons\cite{yang2016nontrivial,tang2020thermal}. However, recent experiments showed that the thermal conductivity of \textit{n}-type GaN is almost a constant within the concentration range of $10^{17} \, \text{to} \, 10^{19} \, \text{cm}^{-3}$\cite{beechem2016size,simon2014thermal}. On the theoretical side, the lattice thermal conductivity of GaN matches experimental results better when accounting for four-phonon scattering\cite{yang2019stronger}. It should be noted that electron-phonon interactions are not included in Ref. \cite{yang2019stronger}\@. Therefore, the effect of electron-phonon interactions on the lattice thermal conductivity of GaN is still under debate.

In this work, the lattice thermal conductivities of \textit{n}-type and \textit{p}-type wurtzite GaN (space group $\mathit{P}{6}_{3}\mathit{mc}$, No. 186) under different carrier concentrations are investigated through mode-level first-principles calculations. It is found that the electron-phonon interactions have weak effects on the lattice thermal conductivity of \textit{n}-type GaN even at an ultra-high electron concentration. The impact of the Fröhlich interaction on lattice thermal conductivity can be ignored, which contradicts the previous understanding on the thermal transport in doped GaN, whereas there is a significant reduction in the lattice thermal conductivity of \textit{p}-type GaN at high hole concentrations. We provide a comprehensive analysis of the contribution terms ascribed to electron-phonon interactions, including the electron density of states (DOS), Fermi surface nesting function, and electron-phonon matrix elements.

\section{THEORY AND METHODS}
Combining the linearized phonon BTE and Fourier's law, the lattice thermal conductivity (${\kappa}_{\text{lat}}$) can be expressed as\cite{broido2007intrinsic},
\begin{equation}
\begin{split}
    \kappa_{\text{lat}, \alpha \beta} &= \sum_{\lambda} c_{v, \lambda} v_{\lambda, \alpha} v_{\lambda, \beta} \tau_{\lambda} \\
    &= \frac{1}{\Omega} \sum_{\lambda} \hbar \omega_{\lambda} \frac{\partial n_{\lambda}}{\partial T} v_{\lambda, \alpha} v_{\lambda, \beta} \tau_{\lambda},
\end{split}
\label{SE1}
\end{equation}
where $\alpha$ and $\beta$ are the Cartesian coordinates, $\lambda \equiv(\mathbf{q}, \nu)$ denotes the phonon mode with wave vector $\mathbf{q}$ and phonon polarization $\nu$, $c_{v, \lambda}$ is the phonon specific heat capacity, $v_{\lambda}$ is the phonon group velocity, and $\tau_{\lambda}$ is the phonon relaxation time. ${\Omega}$ is the volume of the primitive cell, $\hbar$ is the reduced Planck’s constant, $\omega_{\lambda}$ is the phonon frequency, and $n_{\lambda}$ is the Bose-Einstein distribution at temperature ${T}$.

The essential step is to obtain $\tau_{\lambda}$, which is associated with several scattering processes. The effective phonon scattering rates can be obtained using Matthiessen’s rule: $1/\tau_{\lambda}$=$1/\tau_{\lambda}^{\mathrm{ph}-\mathrm{ph}}$+$1/\tau_{\lambda}^{\mathrm{ph}-\text{iso}}$+$1/\tau_{\lambda}^{\mathrm{ph}-\text{el}}$, where $1/\tau_{\lambda}^{\mathrm{ph}-\mathrm{ph}}$ is the phonon-phonon scattering rates, $1/\tau_{\lambda}^{\mathrm{ph}-\text{iso}}$ is the phonon-isotope scattering rates, and $1/\tau_{\lambda}^{\mathrm{ph}-\text{el}}$ is the phonon-electron scattering rates.

According to Fermi's golden rule, the phonon-phonon scattering rates can be expressed as\cite{albers1976normal},
\begin{equation}
\begin{aligned}
    \frac{1}{\tau_{\lambda}^{\mathrm{ph}-\mathrm{ph}}} &= 2 \pi \sum_{\lambda_{1} \lambda_{2}}\left|V_{\lambda \lambda_{1} \lambda_{2}}\right|^{2} \\
    &\times \left[\frac{1}{2}\left(1+n_{\lambda_{1}}^{0}+n_{\lambda_{2}}^{0}\right) \delta\left(\omega_{\lambda}-\omega_{\lambda_{1}}-\omega_{\lambda_{2}}\right) \right. \\
    &\quad + \left. \left(n_{\lambda_{1}}^{0}-n_{\lambda_{2}}^{0}\right) \delta\left(\omega_{\lambda}+\omega_{\lambda_{1}}-\omega_{\lambda_{2}}\right)\right],
\end{aligned}
\label{SE3}
\end{equation}
where $V_{\lambda \lambda_{1} \lambda_{2}}$ denote the three-phonon scattering matrix element. $\delta$ is the Dirac delta function which ensures the conservation of energy during the scattering processes. Recent studies have demonstrated that the four-phonon scattering has strong effects on lattice thermal conductivity at room temperature\cite{zhao2021lattice,zhao2020anomalous,zhao2020quartic,yue2023microscopic,feng2017four}. To evaluate the four-phonon scattering effects on $\kappa_{\text {lat}}$ of GaN, we calculated the weighted phase space of three-phonon and four-phonon scattering. As shown in [Fig. S1, Supplemental Material (see also references \cite{broyden1970convergence,goldfarb1970family,shanno1970conditioning,straumanis1952lattice,schulz1977crystal,ruf2001phonon} therein)], the weighted phase space of four-phonon scattering is significantly smaller than that for three-phonon scattering. Therefore, the four-phonon scattering has negligible effects on the lattice thermal conductivity of GaN at room temperature. The phonon-isotope scattering rates are evaluated by the Tamura theory and the details can be found in Ref. \cite{tamura1983isotope}\@. The phonon-electron scattering rates are related to the imaginary part of the phonon self-energy, which can be expressed as\cite{ponce2016epw},
\begin{equation}
\begin{aligned}
    \frac{1}{\tau_{\lambda}^{\text {ph-el}}} &= -\frac{2 \pi}{\hbar} \sum_{m n, \mathbf{k}}\left|g_{m n}^{v}(\mathbf{k}, \mathbf{q})\right|^{2} \\
    &\quad \times \left(f_{n \mathbf{k}}-f_{m \mathbf{k}+\mathbf{q}}\right) \delta\left(\varepsilon_{m \mathbf{k}+\mathbf{q}}-\varepsilon_{n \mathbf{q}}-\hbar \omega_{\lambda}\right),
\end{aligned}
\label{SE4}
\end{equation}
where $g_{m n, v}(\mathbf{k}, \mathbf{q})$ is the electron-phonon matrix element, which quantifies probability amplitude for scattering between the electronic state $n \mathbf{k}$ and $m \mathbf{k}+\mathbf{q}$, $f$ is the Fermi-Dirac distribution function, $\varepsilon$ is the electron energy, and $\varepsilon_{F}$ is the Fermi energy.

The first-principles calculations are performed using the Quantum Espresso package\cite{giannozzi2009quantum}. The electron exchange-correlation functional is treated by the generalized gradient approximation of Perdew–Burke–Ernzerhof (PBE)\cite{perdew2008restoring} and optimized fully relativistic norm-conserving pseudopotentials\cite{hamann2013optimized} from PseudoDojo\cite{van2018pseudodojo}.  The kinetic energy cutoff for plane waves is set to be 100 Ry, and the convergence of electron energy is $10^{-10}$ Ry. It was shown that the spin-orbit coupling (SOC) has significant effects on the valance band structure in Ref. \cite{ponce2019route}\@. Therefore, the SOC is also included in our electronic band structure calculations. The harmonic and cubic force constants are calculated from density-functional perturbation theory. The electron-phonon interactions are first calculated under coarse $\mathbf{k/q}$ meshes and then interpolated to dense meshes with the Wannier interpolation technique\cite{marzari2012maximally}. The convergence of phonon-electron scattering rates with respect to the \textbf{k}-point mesh is verified in [Fig. S2, Supplemental Material].  The in-house modified D3Q package\cite{paulatto2013anharmonic} is employed to calculate the $\kappa_{\text {lat}}$, incorporating the phonon-electron scattering using the iterative calculation scheme\cite{li2022anomalously,li2020anomalous,sun2023light}. The rigid shift of the Fermi energy is utilized to imitate the change of carrier concentration\cite{li2019resolving}. More details about the first-principles calculations are provided in the Supplemental Material.

\section{RESULTS AND DISCUSSIONS}
\begin{figure}[H]
    \includegraphics[width=1\columnwidth]{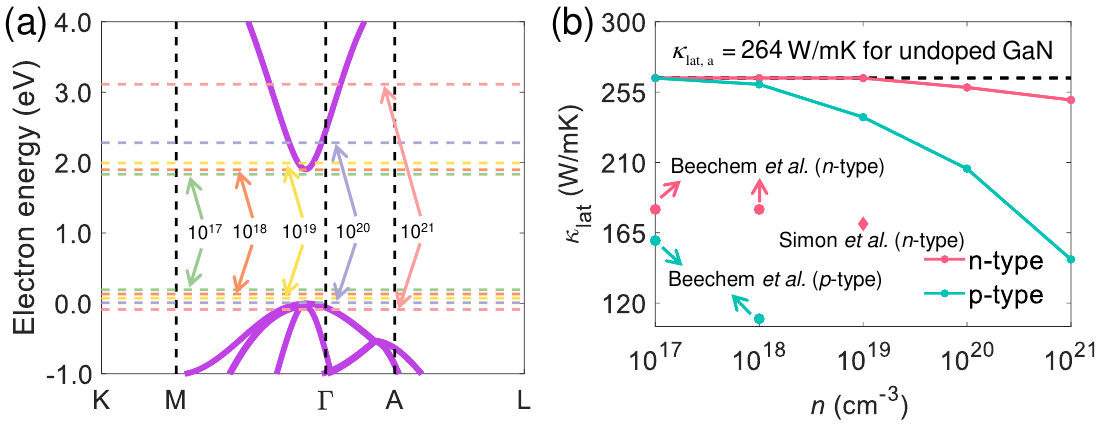}
    \caption {(a) Band structures of GaN along the high-symmetry paths. The horizontal lines are Fermi energy related to the carrier concentrations of $10^{17}$ (green), $10^{18}$ (orange), $10^{19}$ (yellow), $10^{20}$ (light purple), and $10^{21}$ (light pink) $\text{cm}^{-3}$ at room temperature. The electron energy is normalized to the VBM. (b) $\kappa_{1\text{at},a}$ as a function of carrier concentration at room temperature for undoped, \textit{n}-type, and \textit{p}-type GaN. The scatters are experimental results, reported by Beechem \emph{et al}.\cite{beechem2016size} and Simon \emph{et al}.\cite{simon2014thermal}, respectively.}
    \label{fig: Figure 1}
\end{figure}

Figure \ref{fig: Figure 1}(a) shows the electron band structure of GaN that includes the SOC effect. The Fermi energy corresponding to different carrier concentrations is represented by horizontal dashed lines. There is only one conduction band in the vicinity of the conduction band minimum (CBM), while there are multiple valance bands near the valence band maximum (VBM). Note that the heavy-hole and light-hole bands are split along the $\Gamma$-M high-symmetry path of the first Brillouin zone due to the SOC effects, as shown in [Fig. S3(b), Supplemental Material]. This subtle variation, compared to the electronic band structure without SOC, has non-negligible effects on the electron-phonon interactions and the value of in-plane lattice thermal conductivity ($\kappa_{1\text{at},a}$) of {\it p}-type GaN, as shown in Table \ref{tab: Table 1}. Our calculation shows that GaN is an indirect semiconductor with a band gap of 1.86 eV, strongly underestimating the measured value of 3.5 eV\cite{dingle1971absorption,monemar1974fundamental}. This is attributed to the well-known drawback in DFT\cite{chan2010efficient}. However, this discrepancy has no influence on our conclusions on $\kappa_{\text {lat}}$ since the profile of our DFT band structure matches well with the experimental results\cite{rodina2001free} and only the electron modes in the vicinity of the band edges have contributions to electron-phonon interactions. As for the phonon dispersion, our theoretical results agree well with the experimental data [Fig. S4, Supplemental Material], which further confirms the reliability of our first-principles calculations.

\begin{table}[ht]
  \captionsetup{singlelinecheck=false, justification=raggedright}
  \caption{$\kappa_{1\text{at},a}$ of \textit{n}-type and \textit{p}-type GaN with and without SOC/Polar effects. The carrier concentrations are $10^{21} \, \text{cm}^{-3}$ in all the cases.}
  \label{tab: Table 1}
  \begin{tabular}{lcccc}
    \toprule
    $\kappa_{1\text{at},a}$ (300K) & W/ SOC & Wo/ SOC & W/ Polar & Wo/ Polar \\
    \midrule
    \textit{n}-type & 250.34 & 249.59 & 250.34 & 246.51 \\
    \textit{p}-type & 147.85 & 125.51 & 147.85 & 146.20 \\
    \bottomrule
  \end{tabular}
\end{table}

The $\kappa_{1\text{at},a}$ as a function of carrier concentrations ($10^{17}-10^{21} \, \text{cm}^{-3}$) for undoped, \textit{n}-type, and \textit{p}-type GaN are shown in Fig. \ref{fig: Figure 1}(b). The $\kappa_{1\text{at},a}$ at room temperature for undoped GaN, namely without phonon-electron scattering, is 264 W/mK, which is in good agreement with experimental\cite{jezowski2003thermal,slack2002some} and former theoretical results\cite{lindsay2012thermal}. With phonon-electron scattering included, the $\kappa_{1\text{at},a}$ falls within the range of 250-258 W/mK for \textit{n}-type GaN, which is quite close to the value of the undoped case. On the contrary, $\kappa_{1\text{at},a}$ is dramatically reduced to 228 W/mK and 148 W/mK at the hole concentrations of $10^{19} \, \text{cm}^{-3}$ and $10^{21} \, \text{cm}^{-3}$, respectively. A similar variation trend is also observed in time-domain thermoreflectance (TDTR) measurement\cite{beechem2016size,simon2014thermal}, where the $\kappa_{\text {lat}}$ of \textit{n}-type GaN is almost a constant within the doping concentration range of $10^{17}$ to $10^{19} \, \text{cm}^{-3}$. However, the $\kappa_{\text {lat}}$ of \textit{p}-type GaN exhibits a noticeable decrease. The quantitative differences between our work and Refs. \cite{beechem2016size,simon2014thermal} may be attributed to experimental sample sizes, impurities, or defects, which are not considered in our calculations. Recently, Pang \emph{et al}. found that phonon-defect scattering can also have significant effects on the lattice thermal conductivity of 3C-SiC with B doping\cite{pang2024thermal}. Here we mainly focus on the impact of electron-phonon interactions on lattice thermal conductivity. The phonon-defect scattering effects on the lattice thermal conductivity of GaN are out of the perspective of the present work and will be further discussed in future work.

 \begin{figure}[H]
    \centering
    \includegraphics[width=1\columnwidth]{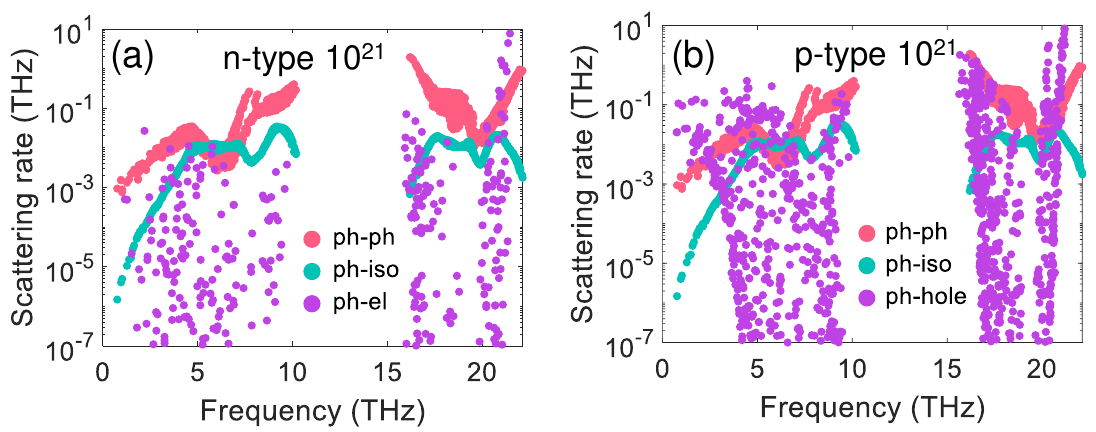}
    \caption {$1 / \tau_{\lambda}^{\mathrm{ph}-\text{ph}}$, $1 / \tau_{\lambda}^{\mathrm{ph}-\text{iso}}$, (a) $1 / \tau_{\lambda}^{\mathrm{ph}-\text{el}}$, and (b) $1 / \tau_{\lambda}^{\mathrm{ph}-\text{hole}}$ at room temperature with carrier concentration of $10^{21} \, \text{cm}^{-3}$.}
    \label{fig: Figure 2}
\end{figure}

\begin{figure*}[hbpt]
    \centering
    \includegraphics[width=2.00\columnwidth]{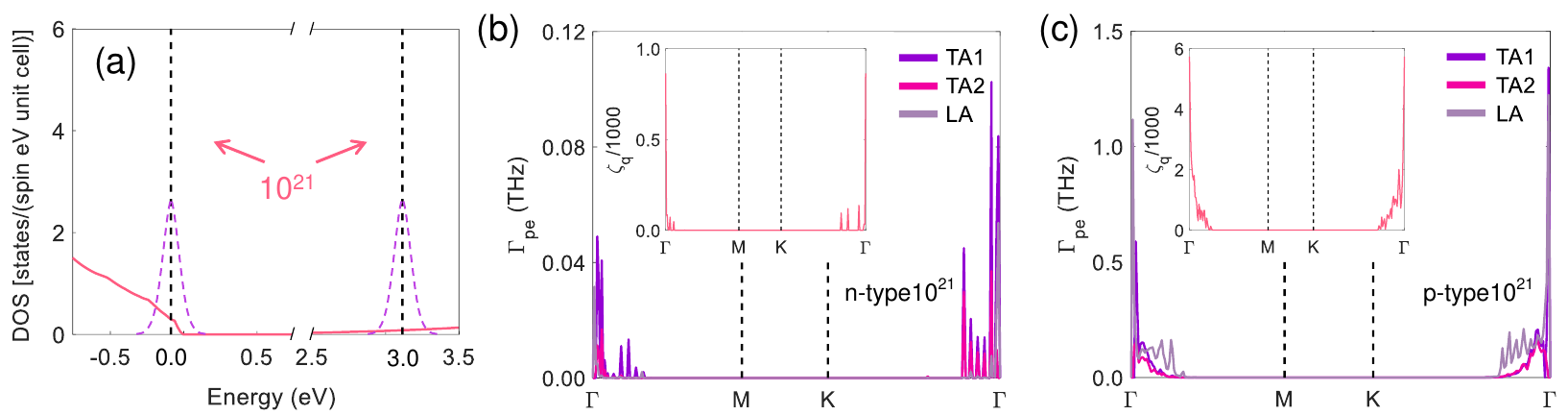}
    \caption{\RaggedRight(a) Electron DOS near the valence- and conduction-band edges. The purple dotted curve represents the Fermi window. The electron energy is normalized to the VBM. The position of the Fermi energy for electron and hole concentrations of $10^{21} \, \text{cm}^{-3}$ is indicated with black dashed lines. Room-temperature phonon linewidth $\Gamma_{\mathrm{pe}}$ of TA1, TA2, and LA along the high-symmetry path due to (b) phonon-electron and (c) phonon-hole scattering with carrier concentration of $10^{21} \, \text{cm}^{-3}$. The Fermi surface nesting function is inserted to the top left in (b) and (c).}
    \label{fig: Figure 3}
\end{figure*}

\begin{figure*}[hbpt]
    \centering
    \includegraphics[width=2.00\columnwidth]{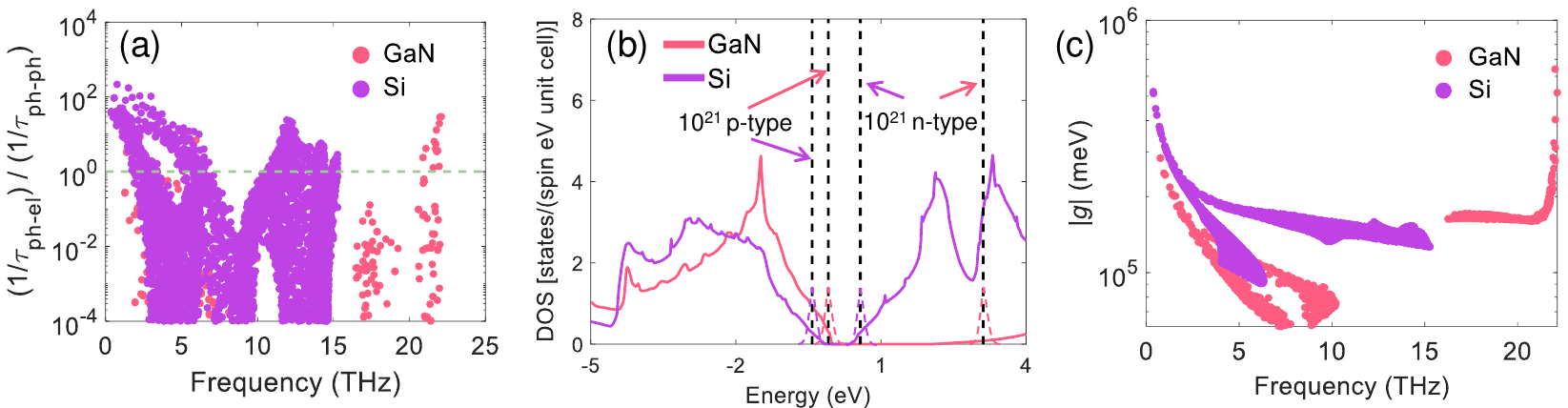}
    \caption {\RaggedRight(a) The ratio of $1 / \tau_{\lambda}^{\mathrm{ph}-\text{el}}$ to $1 / \tau_{\lambda}^{\mathrm{ph}-\text{ph}}$ with $10^{21} \, \text{cm}^{-3}$ electron concentration of GaN and Si, which is used to identify the dominant scattering term in phonon thermal transport. The horizon green line marks the equal importance of $1 / \tau_{\lambda}^{\mathrm{ph}-\text{el}}$ and $1 / \tau_{\lambda}^{\mathrm{ph}-\text{ph}}$. (b) The DOS for GaN and Si. The positions of the Fermi energy for the electron and hole concentrations of $10^{21} \, \text{cm}^{-3}$ are indicated with black dashed lines for GaN and Si. The purple and red dotted curves represent Fermi windows. (c) Absolute value of electron-phonon matrix elements $|\textit{g}|$ of GaN and Si.}
    \label{fig: Figure 4}
\end{figure*}

The effects of electron-phonon interactions on $\kappa_{\text {lat}}$ are significantly different from the observations in Si\cite{liao2015significant} and 3C-SiC\cite{wang2017strong}. Both electron and hole have strong interactions with phonons, resulting in a substantial decrease in $\kappa_{\text {lat}}$\cite{liao2015significant}. Since the $1 / \tau_{\lambda}^{\mathrm{ph}-\text{ph}}$ for phonon frequencies of GaN below 10 THz is smaller than that in Si [Fig. S5, Supplemental Material], the nearly flat $\kappa_{\text {lat}}$ with different concentrations of \textit{n}-type GaN is anomalous. To reveal the underlying mechanisms, the mode-level phonon-electron and phonon-hole scattering rates at the carrier concentration of $10^{21} \, \mathrm{cm}^{-3}$ are calculated, as shown in Figs. \ref{fig: Figure 2}(a) and (b), respectively. The scattering rates at other carrier concentrations are also provided in [Fig. S6, Supplemental Material]. Remarkably, $1 / \tau_{\lambda}^{\mathrm{ph}-\text{el}}$ is significantly lower than $1 / \tau_{\lambda}^{\mathrm{ph}-\text{ph}}$ for low-frequency phonons (below 10THz), which have the main contribution to $\kappa_{\text {lat}}$\cite{wu2016thermal,garg2018spectral}. In contrast, $1 / \tau_{\lambda}^{\mathrm{ph}-\text{hole}}$ is significantly larger than $1 / \tau_{\lambda}^{\mathrm{ph}-\text{ph}}$ within the same frequency range. Note that $1 / \tau_{\lambda}^{\mathrm{ph}-\text{el}}$ and $1 / \tau_{\lambda}^{\mathrm{ph}-\text{hole}}$ for high-frequency optical phonons (above 15 THz) become even larger due to the Fröhlich interaction in polar materials\cite{sheih1995electron}. This phenomenon is also observed in Ref. \cite{yang2016nontrivial}\@. However, optical phonon modes have an ignorable contribution to the $\kappa_{\text {lat}}$ of GaN\cite{wu2016thermal}. Therefore, the strong Fröhlich interaction does not lead to a large reduction in $\kappa_{\text {lat}}$ of GaN. This hypothesis is further verified by the calculated $\kappa_{1\text{at},a}$ with/without considering the Fröhlich interaction, as shown in Table \ref{tab: Table 1}.

The distinctions in $1 / \tau_{\lambda}^{\mathrm{ph}-\text{el}}$ and $1 / \tau_{\lambda}^{\mathrm{ph}-\text{hole}}$ at the same concentration are primarily attributed to the distinctions in electron DOS within the Fermi window regarding $n$-type and $p$-type GaN. GaN becomes a degenerate semiconductor when the carrier concentration reaches $10^{21} \, \text{cm}^{-3}$ for both \textit{n}-type and \textit{p}-type. Therefore, the Fermi energy moves into the conduction band or the valence band. As shown in Fig. \ref{fig: Figure 3}(a), the DOS for holes is notably larger than that for electrons at $10^{21} \, \text{cm}^{-3}$. It should be noted that the DOS exhibits a pronounced asymmetric profile. As shown in [Fig. S7, Supplemental Material], the DOS in the vicinity of the VBM is primarily contributed by \textit{p} orbitals, which exhibit a triple degeneracy and minimal band dispersion. Conversely, the vicinity of the CBM is primarily contributed by \textit{s} orbitals, resulting in nondegenerate and dispersive electron bands\cite{yoodee1984effects,reynolds1999valence,walsh2008origins}. According to Eq. (\ref{SE4}) and its variant based on the double-delta approximation\cite{li2020anomalous,wang2016first,li2021thermal}, the DOS determines the Fermi surface nesting function ($\zeta_{\mathbf{q}}$) to a large extent, which is expressed as
\begin{equation}
    \zeta_{\mathbf{q}}=\sum \delta\left(\varepsilon_{n \mathbf{k}}-\varepsilon_{F}\right) \delta\left(\varepsilon_{m \mathbf{k}+\mathbf{q}}-\varepsilon_{F}\right).
    \label{SE5}
\end{equation}
$\zeta_{\mathbf{q}}$ quantifies the phonon-electron scattering channels\cite{li2018fermi}. The phonon linewidth with respect to the phonon-electron scattering ($\Gamma_{\mathrm{pe}}=1 / 2 \tau_{\lambda}^{\mathrm{ph}-\mathrm{el}}$) and $\zeta_{\mathbf{q}}$ of the two transverse acoustic phonon branches (TA1 and TA2), as well as the longitudinal acoustic phonon branch (LA) are shown in Figs. \ref{fig: Figure 3}(b) and (c). The $\zeta_{\mathbf{q}}$ in the vicinity of the $\Gamma$ point is quite large due to the collinearity of electron group velocities $\mathbf{v}_{\mathbf{k}}$ and $\mathbf{v}_{\mathbf{k}}+\mathbf{q}$\cite{kasinathan2006superconductivity}. The $\zeta_{\mathbf{q}}$ of \textit{n}-type GaN is much lower than that of \textit{p}-type GaN, which results in its relatively smaller $\Gamma_{\mathrm{pe}}$. Therefore, the electron-phonon interactions have weak effects on the $\kappa_{\text {lat}}$ of \textit{n}-type GaN, while they have strong effects on that of \textit{p}-type GaN. This conclusion can be further extended to other wide-bandgap semiconductors, like $\beta$-Ga$_2$O$_3$, AlN, and ZnO\cite{curtarolo2012aflow}, whose electron DOS is relatively larger than the hole DOS.

We further compare the $\kappa_{\text {lat}}$ of Si and GaN for both \textit{n}-type and \textit{p}-type at the carrier concentration of $10^{21} \, \text{cm}^{-3}$. The $\kappa_{\text {lat}}$ of \textit{n}-type Si is reduced by $\sim37\%$\cite{liao2015significant} due to phonon-electron scattering, while the \textit{n}-type GaN is reduced only by 2.85\% in our calculation. As shown in Fig. \ref{fig: Figure 4}(a), $1 / \tau_{\lambda}^{\mathrm{ph}-\text{el}}$ is larger than $1 / \tau_{\lambda}^{\mathrm{ph-ph}}$ for plenty of low-frequency phonons in Si. As a contrast, fewer phonon modes have larger $1 / \tau_{\lambda}^{\mathrm{ph}-\text{el}}$ than $1 / \tau_{\lambda}^{\mathrm{ph}-\text{ph}}$ in GaN. This discrepancy can be attributed to differences in electron DOS. As shown in Fig. \ref{fig: Figure 4}(b), the electron DOS of Si is much larger than that of GaN at the electron concentration of $10^{21} \, \text{cm}^{-3}$. Note that the hole DOS of GaN is close to that of Si at the hole concentration of $10^{21} \, \text{cm}^{-3}$. Nevertheless, the $\kappa_{\text {lat}}$ of Si is decreased by 45\%, while that of GaN is decreased by a smaller value of 29\%. As shown in [Fig. S8, Supplemental Material], the $\zeta_{\mathbf{q}}$ of \textit{p}-type GaN is much larger than that of \textit{p}-type Si, which cannot interpret the discrepancy in their $\kappa_{\text {lat}}$. According to  Eq. (\ref{SE4}), the $1 / \tau_{\lambda}^{\mathrm{ph}-\text{el}}$ is also related to electron-phonon matrix elements $\textit{g}$, which quantifies the coupling strength between phonon modes and electron states,
\begin{equation}
    g_{m n}^{v}(\mathbf{k}, \mathbf{q})=\sqrt{\frac{\hbar}{2 \omega_{\lambda}}}\left\langle\psi_{m \mathbf{k}+\mathbf{q}}\left|\partial_{\lambda} V\right| \psi_{n \mathbf{k}}\right\rangle,
    \label{SE5}
\end{equation}
with $\psi$ the ground-state Bloch wave function and $\partial_{\lambda} V$ the first-order derivative of the Kohn-Sham potential with respect to the atomic displacement. As shown in Fig. \ref{fig: Figure 4}(c), the magnitude of $|\textit{g}|$ for Si is larger than that for GaN, resulting in the smaller $\kappa_{\text {lat}}$ for Si. Recent studies have demonstrated that a 2\% biaxial tensile strain can increase the hole mobility of GaN by 230\%\cite{ponce2019hole,ponce2019route}, which indicates great promise for the application of \textit{p}-type GaN in power electronics. However, our findings reveal that electron-phonon interactions can significantly limit $\kappa_{\text {lat}}$ of \textit{p}-type GaN. Therefore, it is not only urgent to synthesize high-quality \textit{p}-type GaN samples but also important to overcome the challenges in the thermal management of power electronics based on \textit{p}-type GaN in the future.

\section{CONCLUSIONS}
In summary, the lattice thermal conductivities of \textit{n}-type and \textit{p}-type GaN are investigated through mode-level first-principles calculations. The effects of electron-phonon interactions on the lattice thermal conductivity are shown to be weak in \textit{n}-type GaN, even at an ultra-high electron concentration of $10^{21} \, \text{cm}^{-3}$. Intriguingly, our findings indicate that the Fröhlich interaction has an ignorable influence on the lattice thermal conductivity. The weak phonon-electron scattering is attributed to the limited scattering channels, which is reflected by the Fermi surface nesting function. In addition, our study reveals that the electron-phonon interactions significantly limit the lattice thermal conductivities of \textit{p}-type GaN and Si. Importantly, it is the electron-phonon matrix elements, rather than the Fermi surface nesting function, that are ascribed to the relatively larger reduction in the thermal conductivity of \textit{p}-type Si compared to \textit{p}-type GaN. 

\begin{center}
    \vspace{-1em}
    \textbf{ACKNOWLEDGEMENTS}
\end{center}
S.L. was supported by the National Natural Science Foundation of China (Grant No. 12304039), the Shanghai Municipal Natural Science Foundation (Grant No. 22YF1400100), the Fundamental Research Funds for the Central Universities (Grant No. 2232022D-22), and the startup funding for youth faculty from the College of Mechanical Engineering of Donghua University. X.L. was supported by the Shanghai Municipal Natural Science Foundation (Grant No. 21TS1401500) and the National Natural Science Foundation of China (Grants No. 52150610495 and No. 12374027).

\bibliography{bibliography}

\providecommand{\noopsort}[1]{}\providecommand{\singleletter}[1]{#1}%
\begin{thebibliography}{64}%
\makeatletter
\providecommand \@ifxundefined [1]{%
 \@ifx{#1\undefined}
}%
\providecommand \@ifnum [1]{%
 \ifnum #1\expandafter \@firstoftwo
 \else \expandafter \@secondoftwo
 \fi
}%
\providecommand \@ifx [1]{%
 \ifx #1\expandafter \@firstoftwo
 \else \expandafter \@secondoftwo
 \fi
}%
\providecommand \natexlab [1]{#1}%
\providecommand \enquote  [1]{``#1''}%
\providecommand \bibnamefont  [1]{#1}%
\providecommand \bibfnamefont [1]{#1}%
\providecommand \citenamefont [1]{#1}%
\providecommand \href@noop [0]{\@secondoftwo}%
\providecommand \href [0]{\begingroup \@sanitize@url \@href}%
\providecommand \@href[1]{\@@startlink{#1}\@@href}%
\providecommand \@@href[1]{\endgroup#1\@@endlink}%
\providecommand \@sanitize@url [0]{\catcode `\\12\catcode `\$12\catcode `\&12\catcode `\#12\catcode `\^12\catcode `\_12\catcode `\%12\relax}%
\providecommand \@@startlink[1]{}%
\providecommand \@@endlink[0]{}%
\providecommand \url  [0]{\begingroup\@sanitize@url \@url }%
\providecommand \@url [1]{\endgroup\@href {#1}{\urlprefix }}%
\providecommand \urlprefix  [0]{URL }%
\providecommand \Eprint [0]{\href }%
\providecommand \doibase [0]{https://doi.org/}%
\providecommand \selectlanguage [0]{\@gobble}%
\providecommand \bibinfo  [0]{\@secondoftwo}%
\providecommand \bibfield  [0]{\@secondoftwo}%
\providecommand \translation [1]{[#1]}%
\providecommand \BibitemOpen [0]{}%
\providecommand \bibitemStop [0]{}%
\providecommand \bibitemNoStop [0]{.\EOS\space}%
\providecommand \EOS [0]{\spacefactor3000\relax}%
\providecommand \BibitemShut  [1]{\csname bibitem#1\endcsname}%
\let\auto@bib@innerbib\@empty
\bibitem [{\citenamefont {Roccaforte}\ \emph {et~al.}(2018)\citenamefont {Roccaforte}, \citenamefont {Fiorenza}, \citenamefont {Greco}, \citenamefont {Nigro}, \citenamefont {Giannazzo}, \citenamefont {Iucolano},\ and\ \citenamefont {Saggio}}]{roccaforte2018emerging}%
  \BibitemOpen
  \bibfield  {author} {\bibinfo {author} {\bibfnamefont {F.}~\bibnamefont {Roccaforte}}, \bibinfo {author} {\bibfnamefont {P.}~\bibnamefont {Fiorenza}}, \bibinfo {author} {\bibfnamefont {G.}~\bibnamefont {Greco}}, \bibinfo {author} {\bibfnamefont {R.~L.}\ \bibnamefont {Nigro}}, \bibinfo {author} {\bibfnamefont {F.}~\bibnamefont {Giannazzo}}, \bibinfo {author} {\bibfnamefont {F.}~\bibnamefont {Iucolano}},\ and\ \bibinfo {author} {\bibfnamefont {M.}~\bibnamefont {Saggio}},\ }\bibfield  {title} {\bibinfo {title} {Emerging trends in wide band gap semiconductors ({S}i{C} and {G}a{N}) technology for power devices},\ }\href@noop {} {\bibfield  {journal} {\bibinfo  {journal} {Microelectron. Eng.}\ }\textbf {\bibinfo {volume} {187}},\ \bibinfo {pages} {66} (\bibinfo {year} {2018})}\BibitemShut {NoStop}%
\bibitem [{\citenamefont {Chung}\ \emph {et~al.}(2010)\citenamefont {Chung}, \citenamefont {Lee},\ and\ \citenamefont {Yi}}]{chung2010transferable}%
  \BibitemOpen
  \bibfield  {author} {\bibinfo {author} {\bibfnamefont {K.}~\bibnamefont {Chung}}, \bibinfo {author} {\bibfnamefont {C.-H.}\ \bibnamefont {Lee}},\ and\ \bibinfo {author} {\bibfnamefont {G.-C.}\ \bibnamefont {Yi}},\ }\bibfield  {title} {\bibinfo {title} {Transferable {G}a{N} layers grown on {Z}n{O}-coated graphene layers for optoelectronic devices},\ }\href@noop {} {\bibfield  {journal} {\bibinfo  {journal} {Science}\ }\textbf {\bibinfo {volume} {330}},\ \bibinfo {pages} {655} (\bibinfo {year} {2010})}\BibitemShut {NoStop}%
\bibitem [{\citenamefont {Korotkov}\ \emph {et~al.}(2001)\citenamefont {Korotkov}, \citenamefont {Gregie},\ and\ \citenamefont {Wessels}}]{korotkov2001electrical}%
  \BibitemOpen
  \bibfield  {author} {\bibinfo {author} {\bibfnamefont {R.}~\bibnamefont {Korotkov}}, \bibinfo {author} {\bibfnamefont {J.}~\bibnamefont {Gregie}},\ and\ \bibinfo {author} {\bibfnamefont {B.~W.}\ \bibnamefont {Wessels}},\ }\bibfield  {title} {\bibinfo {title} {Electrical properties of p-type {G}a{N}: Mg codoped with oxygen},\ }\href@noop {} {\bibfield  {journal} {\bibinfo  {journal} {Appl. Phys. Lett.}\ }\textbf {\bibinfo {volume} {78}},\ \bibinfo {pages} {222} (\bibinfo {year} {2001})}\BibitemShut {NoStop}%
\bibitem [{\citenamefont {Konczewicz}\ \emph {et~al.}(2022)\citenamefont {Konczewicz}, \citenamefont {Litwin-Staszewska}, \citenamefont {Zajac}, \citenamefont {Turski}, \citenamefont {Bockowski}, \citenamefont {Schiavon}, \citenamefont {Chlipa{\l}a}, \citenamefont {Iwinska}, \citenamefont {Nita}, \citenamefont {Juillaguet} \emph {et~al.}}]{konczewicz2022electrical}%
  \BibitemOpen
  \bibfield  {author} {\bibinfo {author} {\bibfnamefont {L.}~\bibnamefont {Konczewicz}}, \bibinfo {author} {\bibfnamefont {E.}~\bibnamefont {Litwin-Staszewska}}, \bibinfo {author} {\bibfnamefont {M.}~\bibnamefont {Zajac}}, \bibinfo {author} {\bibfnamefont {H.}~\bibnamefont {Turski}}, \bibinfo {author} {\bibfnamefont {M.}~\bibnamefont {Bockowski}}, \bibinfo {author} {\bibfnamefont {D.}~\bibnamefont {Schiavon}}, \bibinfo {author} {\bibfnamefont {M.}~\bibnamefont {Chlipa{\l}a}}, \bibinfo {author} {\bibfnamefont {M.}~\bibnamefont {Iwinska}}, \bibinfo {author} {\bibfnamefont {P.}~\bibnamefont {Nita}}, \bibinfo {author} {\bibfnamefont {S.}~\bibnamefont {Juillaguet}}, \emph {et~al.},\ }\bibfield  {title} {\bibinfo {title} {Electrical transport properties of highly doped n-type {G}a{N} materials},\ }\href@noop {} {\bibfield  {journal} {\bibinfo  {journal} {Semicond. Sci. Technol.}\ }\textbf {\bibinfo {volume} {37}},\ \bibinfo {pages} {055012} (\bibinfo {year} {2022})}\BibitemShut {NoStop}%
\bibitem [{\citenamefont {Warzoha}\ \emph {et~al.}(2021)\citenamefont {Warzoha}, \citenamefont {Wilson}, \citenamefont {Donovan}, \citenamefont {Donmezer}, \citenamefont {Giri}, \citenamefont {Hopkins}, \citenamefont {Choi}, \citenamefont {Pahinkar}, \citenamefont {Shi}, \citenamefont {Graham} \emph {et~al.}}]{warzoha2021applications}%
  \BibitemOpen
  \bibfield  {author} {\bibinfo {author} {\bibfnamefont {R.~J.}\ \bibnamefont {Warzoha}}, \bibinfo {author} {\bibfnamefont {A.~A.}\ \bibnamefont {Wilson}}, \bibinfo {author} {\bibfnamefont {B.~F.}\ \bibnamefont {Donovan}}, \bibinfo {author} {\bibfnamefont {N.}~\bibnamefont {Donmezer}}, \bibinfo {author} {\bibfnamefont {A.}~\bibnamefont {Giri}}, \bibinfo {author} {\bibfnamefont {P.~E.}\ \bibnamefont {Hopkins}}, \bibinfo {author} {\bibfnamefont {S.}~\bibnamefont {Choi}}, \bibinfo {author} {\bibfnamefont {D.}~\bibnamefont {Pahinkar}}, \bibinfo {author} {\bibfnamefont {J.}~\bibnamefont {Shi}}, \bibinfo {author} {\bibfnamefont {S.}~\bibnamefont {Graham}}, \emph {et~al.},\ }\bibfield  {title} {\bibinfo {title} {Applications and impacts of nanoscale thermal transport in electronics packaging},\ }\href@noop {} {\bibfield  {journal} {\bibinfo  {journal} {J. Electron. Packag.}\ }\textbf {\bibinfo {volume} {143}},\ \bibinfo {pages} {020804} (\bibinfo {year} {2021})}\BibitemShut {NoStop}%
\bibitem [{\citenamefont {Tong}\ \emph {et~al.}(2019)\citenamefont {Tong}, \citenamefont {Li}, \citenamefont {Ruan},\ and\ \citenamefont {Bao}}]{tong2019comprehensive}%
  \BibitemOpen
  \bibfield  {author} {\bibinfo {author} {\bibfnamefont {Z.}~\bibnamefont {Tong}}, \bibinfo {author} {\bibfnamefont {S.}~\bibnamefont {Li}}, \bibinfo {author} {\bibfnamefont {X.}~\bibnamefont {Ruan}},\ and\ \bibinfo {author} {\bibfnamefont {H.}~\bibnamefont {Bao}},\ }\bibfield  {title} {\bibinfo {title} {Comprehensive first-principles analysis of phonon thermal conductivity and electron-phonon coupling in different metals},\ }\href@noop {} {\bibfield  {journal} {\bibinfo  {journal} {Phys. Rev. B}\ }\textbf {\bibinfo {volume} {100}},\ \bibinfo {pages} {144306} (\bibinfo {year} {2019})}\BibitemShut {NoStop}%
\bibitem [{\citenamefont {Jain}\ and\ \citenamefont {McGaughey}(2016)}]{jain2016thermal}%
  \BibitemOpen
  \bibfield  {author} {\bibinfo {author} {\bibfnamefont {A.}~\bibnamefont {Jain}}\ and\ \bibinfo {author} {\bibfnamefont {A.~J.}\ \bibnamefont {McGaughey}},\ }\bibfield  {title} {\bibinfo {title} {Thermal transport by phonons and electrons in aluminum, silver, and gold from first principles},\ }\href@noop {} {\bibfield  {journal} {\bibinfo  {journal} {Phys. Rev. B}\ }\textbf {\bibinfo {volume} {93}},\ \bibinfo {pages} {081206} (\bibinfo {year} {2016})}\BibitemShut {NoStop}%
\bibitem [{\citenamefont {Chen}\ \emph {et~al.}(2019)\citenamefont {Chen}, \citenamefont {Ma},\ and\ \citenamefont {Li}}]{chen2019understanding}%
  \BibitemOpen
  \bibfield  {author} {\bibinfo {author} {\bibfnamefont {Y.}~\bibnamefont {Chen}}, \bibinfo {author} {\bibfnamefont {J.}~\bibnamefont {Ma}},\ and\ \bibinfo {author} {\bibfnamefont {W.}~\bibnamefont {Li}},\ }\bibfield  {title} {\bibinfo {title} {Understanding the thermal conductivity and lorenz number in tungsten from first principles},\ }\href@noop {} {\bibfield  {journal} {\bibinfo  {journal} {Phys. Rev. B}\ }\textbf {\bibinfo {volume} {99}},\ \bibinfo {pages} {020305} (\bibinfo {year} {2019})}\BibitemShut {NoStop}%
\bibitem [{\citenamefont {Liao}\ \emph {et~al.}(2015)\citenamefont {Liao}, \citenamefont {Qiu}, \citenamefont {Zhou}, \citenamefont {Huberman}, \citenamefont {Esfarjani},\ and\ \citenamefont {Chen}}]{liao2015significant}%
  \BibitemOpen
  \bibfield  {author} {\bibinfo {author} {\bibfnamefont {B.}~\bibnamefont {Liao}}, \bibinfo {author} {\bibfnamefont {B.}~\bibnamefont {Qiu}}, \bibinfo {author} {\bibfnamefont {J.}~\bibnamefont {Zhou}}, \bibinfo {author} {\bibfnamefont {S.}~\bibnamefont {Huberman}}, \bibinfo {author} {\bibfnamefont {K.}~\bibnamefont {Esfarjani}},\ and\ \bibinfo {author} {\bibfnamefont {G.}~\bibnamefont {Chen}},\ }\bibfield  {title} {\bibinfo {title} {Significant reduction of lattice thermal conductivity by the electron-phonon interaction in silicon with high carrier concentrations: A first-principles study},\ }\href@noop {} {\bibfield  {journal} {\bibinfo  {journal} {Phys. Rev. Lett.}\ }\textbf {\bibinfo {volume} {114}},\ \bibinfo {pages} {115901} (\bibinfo {year} {2015})}\BibitemShut {NoStop}%
\bibitem [{\citenamefont {Liao}\ \emph {et~al.}(2016)\citenamefont {Liao}, \citenamefont {Maznev}, \citenamefont {Nelson},\ and\ \citenamefont {Chen}}]{liao2016photo}%
  \BibitemOpen
  \bibfield  {author} {\bibinfo {author} {\bibfnamefont {B.}~\bibnamefont {Liao}}, \bibinfo {author} {\bibfnamefont {A.}~\bibnamefont {Maznev}}, \bibinfo {author} {\bibfnamefont {K.~A.}\ \bibnamefont {Nelson}},\ and\ \bibinfo {author} {\bibfnamefont {G.}~\bibnamefont {Chen}},\ }\bibfield  {title} {\bibinfo {title} {Photo-excited charge carriers suppress sub-terahertz phonon mode in silicon at room temperature},\ }\href@noop {} {\bibfield  {journal} {\bibinfo  {journal} {Nat. Commun.}\ }\textbf {\bibinfo {volume} {7}},\ \bibinfo {pages} {13174} (\bibinfo {year} {2016})}\BibitemShut {NoStop}%
\bibitem [{\citenamefont {Liu}\ \emph {et~al.}(2020)\citenamefont {Liu}, \citenamefont {Yao}, \citenamefont {Yang}, \citenamefont {Xi},\ and\ \citenamefont {Ke}}]{liu2020strong}%
  \BibitemOpen
  \bibfield  {author} {\bibinfo {author} {\bibfnamefont {C.}~\bibnamefont {Liu}}, \bibinfo {author} {\bibfnamefont {M.}~\bibnamefont {Yao}}, \bibinfo {author} {\bibfnamefont {J.}~\bibnamefont {Yang}}, \bibinfo {author} {\bibfnamefont {J.}~\bibnamefont {Xi}},\ and\ \bibinfo {author} {\bibfnamefont {X.}~\bibnamefont {Ke}},\ }\bibfield  {title} {\bibinfo {title} {Strong electron-phonon interaction induced significant reduction in lattice thermal conductivities for single-layer {M}o{S}$_{2}$ and {P}t{S}{S}e},\ }\href@noop {} {\bibfield  {journal} {\bibinfo  {journal} {Mater. Today Phys.}\ }\textbf {\bibinfo {volume} {15}},\ \bibinfo {pages} {100277} (\bibinfo {year} {2020})}\BibitemShut {NoStop}%
\bibitem [{\citenamefont {Yue}\ \emph {et~al.}(2019)\citenamefont {Yue}, \citenamefont {Yang},\ and\ \citenamefont {Liao}}]{yue2019controlling}%
  \BibitemOpen
  \bibfield  {author} {\bibinfo {author} {\bibfnamefont {S.-Y.}\ \bibnamefont {Yue}}, \bibinfo {author} {\bibfnamefont {R.}~\bibnamefont {Yang}},\ and\ \bibinfo {author} {\bibfnamefont {B.}~\bibnamefont {Liao}},\ }\bibfield  {title} {\bibinfo {title} {Controlling thermal conductivity of two-dimensional materials via externally induced phonon-electron interaction},\ }\href@noop {} {\bibfield  {journal} {\bibinfo  {journal} {Phys. Rev. B}\ }\textbf {\bibinfo {volume} {100}},\ \bibinfo {pages} {115408} (\bibinfo {year} {2019})}\BibitemShut {NoStop}%
\bibitem [{\citenamefont {Lindsay}\ \emph {et~al.}(2012)\citenamefont {Lindsay}, \citenamefont {Broido},\ and\ \citenamefont {Reinecke}}]{lindsay2012thermal}%
  \BibitemOpen
  \bibfield  {author} {\bibinfo {author} {\bibfnamefont {L.}~\bibnamefont {Lindsay}}, \bibinfo {author} {\bibfnamefont {D.}~\bibnamefont {Broido}},\ and\ \bibinfo {author} {\bibfnamefont {T.}~\bibnamefont {Reinecke}},\ }\bibfield  {title} {\bibinfo {title} {Thermal conductivity and large isotope effect in {G}a{N} from first principles},\ }\href@noop {} {\bibfield  {journal} {\bibinfo  {journal} {Phys. Rev. Lett.}\ }\textbf {\bibinfo {volume} {109}},\ \bibinfo {pages} {095901} (\bibinfo {year} {2012})}\BibitemShut {NoStop}%
\bibitem [{\citenamefont {Wang}\ \emph {et~al.}(2017)\citenamefont {Wang}, \citenamefont {Gui}, \citenamefont {Janotti}, \citenamefont {Ni},\ and\ \citenamefont {Karandikar}}]{wang2017strong}%
  \BibitemOpen
  \bibfield  {author} {\bibinfo {author} {\bibfnamefont {T.}~\bibnamefont {Wang}}, \bibinfo {author} {\bibfnamefont {Z.}~\bibnamefont {Gui}}, \bibinfo {author} {\bibfnamefont {A.}~\bibnamefont {Janotti}}, \bibinfo {author} {\bibfnamefont {C.}~\bibnamefont {Ni}},\ and\ \bibinfo {author} {\bibfnamefont {P.}~\bibnamefont {Karandikar}},\ }\bibfield  {title} {\bibinfo {title} {Strong effect of electron-phonon interaction on the lattice thermal conductivity in 3{C}-{S}i{C}},\ }\href@noop {} {\bibfield  {journal} {\bibinfo  {journal} {Phys. Rev. Mater.}\ }\textbf {\bibinfo {volume} {1}},\ \bibinfo {pages} {034601} (\bibinfo {year} {2017})}\BibitemShut {NoStop}%
\bibitem [{\citenamefont {Yang}\ \emph {et~al.}(2016)\citenamefont {Yang}, \citenamefont {Qin},\ and\ \citenamefont {Hu}}]{yang2016nontrivial}%
  \BibitemOpen
  \bibfield  {author} {\bibinfo {author} {\bibfnamefont {J.-Y.}\ \bibnamefont {Yang}}, \bibinfo {author} {\bibfnamefont {G.}~\bibnamefont {Qin}},\ and\ \bibinfo {author} {\bibfnamefont {M.}~\bibnamefont {Hu}},\ }\bibfield  {title} {\bibinfo {title} {Nontrivial contribution of {F}r{\"o}hlich electron-phonon interaction to lattice thermal conductivity of wurtzite {G}a{N}},\ }\href@noop {} {\bibfield  {journal} {\bibinfo  {journal} {Appl. Phys. Lett.}\ }\textbf {\bibinfo {volume} {109}},\ \bibinfo {pages} {242103} (\bibinfo {year} {2016})}\BibitemShut {NoStop}%
\bibitem [{\citenamefont {Tang}\ \emph {et~al.}(2020)\citenamefont {Tang}, \citenamefont {Qin}, \citenamefont {Hu},\ and\ \citenamefont {Cao}}]{tang2020thermal}%
  \BibitemOpen
  \bibfield  {author} {\bibinfo {author} {\bibfnamefont {D.-S.}\ \bibnamefont {Tang}}, \bibinfo {author} {\bibfnamefont {G.-Z.}\ \bibnamefont {Qin}}, \bibinfo {author} {\bibfnamefont {M.}~\bibnamefont {Hu}},\ and\ \bibinfo {author} {\bibfnamefont {B.-Y.}\ \bibnamefont {Cao}},\ }\bibfield  {title} {\bibinfo {title} {Thermal transport properties of {G}a{N} with biaxial strain and electron-phonon coupling},\ }\href@noop {} {\bibfield  {journal} {\bibinfo  {journal} {J. Appl. Phys.}\ }\textbf {\bibinfo {volume} {127}},\ \bibinfo {pages} {035102} (\bibinfo {year} {2020})}\BibitemShut {NoStop}%
\bibitem [{\citenamefont {Beechem}\ \emph {et~al.}(2016)\citenamefont {Beechem}, \citenamefont {McDonald}, \citenamefont {Fuller}, \citenamefont {Talin}, \citenamefont {Rost}, \citenamefont {Maria}, \citenamefont {Gaskins}, \citenamefont {Hopkins},\ and\ \citenamefont {Allerman}}]{beechem2016size}%
  \BibitemOpen
  \bibfield  {author} {\bibinfo {author} {\bibfnamefont {T.~E.}\ \bibnamefont {Beechem}}, \bibinfo {author} {\bibfnamefont {A.~E.}\ \bibnamefont {McDonald}}, \bibinfo {author} {\bibfnamefont {E.~J.}\ \bibnamefont {Fuller}}, \bibinfo {author} {\bibfnamefont {A.~A.}\ \bibnamefont {Talin}}, \bibinfo {author} {\bibfnamefont {C.~M.}\ \bibnamefont {Rost}}, \bibinfo {author} {\bibfnamefont {J.-P.}\ \bibnamefont {Maria}}, \bibinfo {author} {\bibfnamefont {J.~T.}\ \bibnamefont {Gaskins}}, \bibinfo {author} {\bibfnamefont {P.~E.}\ \bibnamefont {Hopkins}},\ and\ \bibinfo {author} {\bibfnamefont {A.~A.}\ \bibnamefont {Allerman}},\ }\bibfield  {title} {\bibinfo {title} {Size dictated thermal conductivity of {G}a{N}},\ }\href@noop {} {\bibfield  {journal} {\bibinfo  {journal} {J. Appl. Phys.}\ }\textbf {\bibinfo {volume} {120}},\ \bibinfo {pages} {095104} (\bibinfo {year} {2016})}\BibitemShut {NoStop}%
\bibitem [{\citenamefont {Simon}\ \emph {et~al.}(2014)\citenamefont {Simon}, \citenamefont {Anaya},\ and\ \citenamefont {Kuball}}]{simon2014thermal}%
  \BibitemOpen
  \bibfield  {author} {\bibinfo {author} {\bibfnamefont {R.~B.}\ \bibnamefont {Simon}}, \bibinfo {author} {\bibfnamefont {J.}~\bibnamefont {Anaya}},\ and\ \bibinfo {author} {\bibfnamefont {M.}~\bibnamefont {Kuball}},\ }\bibfield  {title} {\bibinfo {title} {Thermal conductivity of bulk {G}a{N}—{E}ffects of oxygen, magnesium doping, and strain field compensation},\ }\href@noop {} {\bibfield  {journal} {\bibinfo  {journal} {Appl. Phys. Lett.}\ }\textbf {\bibinfo {volume} {105}},\ \bibinfo {pages} {202105} (\bibinfo {year} {2014})}\BibitemShut {NoStop}%
\bibitem [{\citenamefont {Yang}\ \emph {et~al.}(2019)\citenamefont {Yang}, \citenamefont {Feng}, \citenamefont {Li},\ and\ \citenamefont {Ruan}}]{yang2019stronger}%
  \BibitemOpen
  \bibfield  {author} {\bibinfo {author} {\bibfnamefont {X.}~\bibnamefont {Yang}}, \bibinfo {author} {\bibfnamefont {T.}~\bibnamefont {Feng}}, \bibinfo {author} {\bibfnamefont {J.}~\bibnamefont {Li}},\ and\ \bibinfo {author} {\bibfnamefont {X.}~\bibnamefont {Ruan}},\ }\bibfield  {title} {\bibinfo {title} {Stronger role of four-phonon scattering than three-phonon scattering in thermal conductivity of {III-V} semiconductors at room temperature},\ }\href@noop {} {\bibfield  {journal} {\bibinfo  {journal} {Phys. Rev. B}\ }\textbf {\bibinfo {volume} {100}},\ \bibinfo {pages} {245203} (\bibinfo {year} {2019})}\BibitemShut {NoStop}%
\bibitem [{\citenamefont {Broido}\ \emph {et~al.}(2007)\citenamefont {Broido}, \citenamefont {Malorny}, \citenamefont {Birner}, \citenamefont {Mingo},\ and\ \citenamefont {Stewart}}]{broido2007intrinsic}%
  \BibitemOpen
  \bibfield  {author} {\bibinfo {author} {\bibfnamefont {D.~A.}\ \bibnamefont {Broido}}, \bibinfo {author} {\bibfnamefont {M.}~\bibnamefont {Malorny}}, \bibinfo {author} {\bibfnamefont {G.}~\bibnamefont {Birner}}, \bibinfo {author} {\bibfnamefont {N.}~\bibnamefont {Mingo}},\ and\ \bibinfo {author} {\bibfnamefont {D.}~\bibnamefont {Stewart}},\ }\bibfield  {title} {\bibinfo {title} {Intrinsic lattice thermal conductivity of semiconductors from first principles},\ }\href@noop {} {\bibfield  {journal} {\bibinfo  {journal} {Appl. Phys. Lett.}\ }\textbf {\bibinfo {volume} {91}},\ \bibinfo {pages} {231922} (\bibinfo {year} {2007})}\BibitemShut {NoStop}%
\bibitem [{\citenamefont {Albers}\ \emph {et~al.}(1976)\citenamefont {Albers}, \citenamefont {Bohlin}, \citenamefont {Roy},\ and\ \citenamefont {Wilkins}}]{albers1976normal}%
  \BibitemOpen
  \bibfield  {author} {\bibinfo {author} {\bibfnamefont {R.}~\bibnamefont {Albers}}, \bibinfo {author} {\bibfnamefont {L.}~\bibnamefont {Bohlin}}, \bibinfo {author} {\bibfnamefont {M.}~\bibnamefont {Roy}},\ and\ \bibinfo {author} {\bibfnamefont {J.}~\bibnamefont {Wilkins}},\ }\bibfield  {title} {\bibinfo {title} {Normal and umklapp phonon decay rates due to phonon-phonon and electron-phonon scattering in potassium at low temperatures},\ }\href@noop {} {\bibfield  {journal} {\bibinfo  {journal} {Phys. Rev. B}\ }\textbf {\bibinfo {volume} {13}},\ \bibinfo {pages} {768} (\bibinfo {year} {1976})}\BibitemShut {NoStop}%
\bibitem [{\citenamefont {Zhao}\ \emph {et~al.}(2021)\citenamefont {Zhao}, \citenamefont {Zeng}, \citenamefont {Li}, \citenamefont {Lian}, \citenamefont {Dai}, \citenamefont {Meng},\ and\ \citenamefont {Ni}}]{zhao2021lattice}%
  \BibitemOpen
  \bibfield  {author} {\bibinfo {author} {\bibfnamefont {Y.}~\bibnamefont {Zhao}}, \bibinfo {author} {\bibfnamefont {S.}~\bibnamefont {Zeng}}, \bibinfo {author} {\bibfnamefont {G.}~\bibnamefont {Li}}, \bibinfo {author} {\bibfnamefont {C.}~\bibnamefont {Lian}}, \bibinfo {author} {\bibfnamefont {Z.}~\bibnamefont {Dai}}, \bibinfo {author} {\bibfnamefont {S.}~\bibnamefont {Meng}},\ and\ \bibinfo {author} {\bibfnamefont {J.}~\bibnamefont {Ni}},\ }\bibfield  {title} {\bibinfo {title} {Lattice thermal conductivity including phonon frequency shifts and scattering rates induced by quartic anharmonicity in cubic oxide and fluoride perovskites},\ }\href@noop {} {\bibfield  {journal} {\bibinfo  {journal} {Phys. Rev. B}\ }\textbf {\bibinfo {volume} {104}},\ \bibinfo {pages} {224304} (\bibinfo {year} {2021})}\BibitemShut {NoStop}%
\bibitem [{\citenamefont {Zhao}\ \emph {et~al.}(2020{\natexlab{a}})\citenamefont {Zhao}, \citenamefont {Lian}, \citenamefont {Zeng}, \citenamefont {Dai}, \citenamefont {Meng},\ and\ \citenamefont {Ni}}]{zhao2020anomalous}%
  \BibitemOpen
  \bibfield  {author} {\bibinfo {author} {\bibfnamefont {Y.}~\bibnamefont {Zhao}}, \bibinfo {author} {\bibfnamefont {C.}~\bibnamefont {Lian}}, \bibinfo {author} {\bibfnamefont {S.}~\bibnamefont {Zeng}}, \bibinfo {author} {\bibfnamefont {Z.}~\bibnamefont {Dai}}, \bibinfo {author} {\bibfnamefont {S.}~\bibnamefont {Meng}},\ and\ \bibinfo {author} {\bibfnamefont {J.}~\bibnamefont {Ni}},\ }\bibfield  {title} {\bibinfo {title} {Anomalous electronic and thermoelectric transport properties in cubic {R}b$_{3}${A}u{O} antiperovskite},\ }\href@noop {} {\bibfield  {journal} {\bibinfo  {journal} {Phys. Rev. B}\ }\textbf {\bibinfo {volume} {102}},\ \bibinfo {pages} {094314} (\bibinfo {year} {2020}{\natexlab{a}})}\BibitemShut {NoStop}%
\bibitem [{\citenamefont {Zhao}\ \emph {et~al.}(2020{\natexlab{b}})\citenamefont {Zhao}, \citenamefont {Lian}, \citenamefont {Zeng}, \citenamefont {Dai}, \citenamefont {Meng},\ and\ \citenamefont {Ni}}]{zhao2020quartic}%
  \BibitemOpen
  \bibfield  {author} {\bibinfo {author} {\bibfnamefont {Y.}~\bibnamefont {Zhao}}, \bibinfo {author} {\bibfnamefont {C.}~\bibnamefont {Lian}}, \bibinfo {author} {\bibfnamefont {S.}~\bibnamefont {Zeng}}, \bibinfo {author} {\bibfnamefont {Z.}~\bibnamefont {Dai}}, \bibinfo {author} {\bibfnamefont {S.}~\bibnamefont {Meng}},\ and\ \bibinfo {author} {\bibfnamefont {J.}~\bibnamefont {Ni}},\ }\bibfield  {title} {\bibinfo {title} {Quartic anharmonicity and anomalous thermal conductivity in cubic antiperovskites {A}$_{3}${BO} ({A= K, R}b; {B= B}r, {A}u)},\ }\href@noop {} {\bibfield  {journal} {\bibinfo  {journal} {Phys. Rev. B}\ }\textbf {\bibinfo {volume} {101}},\ \bibinfo {pages} {184303} (\bibinfo {year} {2020}{\natexlab{b}})}\BibitemShut {NoStop}%
\bibitem [{\citenamefont {Yue}\ \emph {et~al.}(2023)\citenamefont {Yue}, \citenamefont {Zhao}, \citenamefont {Ni}, \citenamefont {Meng},\ and\ \citenamefont {Dai}}]{yue2023microscopic}%
  \BibitemOpen
  \bibfield  {author} {\bibinfo {author} {\bibfnamefont {T.}~\bibnamefont {Yue}}, \bibinfo {author} {\bibfnamefont {Y.}~\bibnamefont {Zhao}}, \bibinfo {author} {\bibfnamefont {J.}~\bibnamefont {Ni}}, \bibinfo {author} {\bibfnamefont {S.}~\bibnamefont {Meng}},\ and\ \bibinfo {author} {\bibfnamefont {Z.}~\bibnamefont {Dai}},\ }\bibfield  {title} {\bibinfo {title} {Microscopic mechanism of low lattice thermal conductivity in natural superlattice materials {B}a{XYF} ({X= C}u, {A}g; {Y= S}e, {T}e) including fully quartic anharmonicity},\ }\href@noop {} {\bibfield  {journal} {\bibinfo  {journal} {Phys. Rev. B}\ }\textbf {\bibinfo {volume} {107}},\ \bibinfo {pages} {024301} (\bibinfo {year} {2023})}\BibitemShut {NoStop}%
\bibitem [{\citenamefont {Feng}\ \emph {et~al.}(2017)\citenamefont {Feng}, \citenamefont {Lindsay},\ and\ \citenamefont {Ruan}}]{feng2017four}%
  \BibitemOpen
  \bibfield  {author} {\bibinfo {author} {\bibfnamefont {T.}~\bibnamefont {Feng}}, \bibinfo {author} {\bibfnamefont {L.}~\bibnamefont {Lindsay}},\ and\ \bibinfo {author} {\bibfnamefont {X.}~\bibnamefont {Ruan}},\ }\bibfield  {title} {\bibinfo {title} {Four-phonon scattering significantly reduces intrinsic thermal conductivity of solids},\ }\href@noop {} {\bibfield  {journal} {\bibinfo  {journal} {Phys. Rev. B}\ }\textbf {\bibinfo {volume} {96}},\ \bibinfo {pages} {161201} (\bibinfo {year} {2017})}\BibitemShut {NoStop}%
\bibitem [{\citenamefont {Broyden}(1970)}]{broyden1970convergence}%
  \BibitemOpen
  \bibfield  {author} {\bibinfo {author} {\bibfnamefont {C.~G.}\ \bibnamefont {Broyden}},\ }\bibfield  {title} {\bibinfo {title} {The convergence of a class of double-rank minimization algorithms 1. general considerations},\ }\href@noop {} {\bibfield  {journal} {\bibinfo  {journal} {J. Inst. Math. Its Appl.}\ }\textbf {\bibinfo {volume} {6}},\ \bibinfo {pages} {76} (\bibinfo {year} {1970})}\BibitemShut {NoStop}%
\bibitem [{\citenamefont {Goldfarb}(1970)}]{goldfarb1970family}%
  \BibitemOpen
  \bibfield  {author} {\bibinfo {author} {\bibfnamefont {D.}~\bibnamefont {Goldfarb}},\ }\bibfield  {title} {\bibinfo {title} {A family of variable-metric methods derived by variational means},\ }\href@noop {} {\bibfield  {journal} {\bibinfo  {journal} {Math. Comput.}\ }\textbf {\bibinfo {volume} {24}},\ \bibinfo {pages} {23} (\bibinfo {year} {1970})}\BibitemShut {NoStop}%
\bibitem [{\citenamefont {Shanno}(1970)}]{shanno1970conditioning}%
  \BibitemOpen
  \bibfield  {author} {\bibinfo {author} {\bibfnamefont {D.~F.}\ \bibnamefont {Shanno}},\ }\bibfield  {title} {\bibinfo {title} {Conditioning of quasi-{N}ewton methods for function minimization},\ }\href@noop {} {\bibfield  {journal} {\bibinfo  {journal} {Math. Comput.}\ }\textbf {\bibinfo {volume} {24}},\ \bibinfo {pages} {647} (\bibinfo {year} {1970})}\BibitemShut {NoStop}%
\bibitem [{\citenamefont {Straumanis}\ and\ \citenamefont {Aka}(1952)}]{straumanis1952lattice}%
  \BibitemOpen
  \bibfield  {author} {\bibinfo {author} {\bibfnamefont {M.}~\bibnamefont {Straumanis}}\ and\ \bibinfo {author} {\bibfnamefont {E.}~\bibnamefont {Aka}},\ }\bibfield  {title} {\bibinfo {title} {Lattice parameters, coefficients of thermal expansion, and atomic weights of purest silicon and germanium},\ }\href@noop {} {\bibfield  {journal} {\bibinfo  {journal} {J. Appl. Phys.}\ }\textbf {\bibinfo {volume} {23}},\ \bibinfo {pages} {330} (\bibinfo {year} {1952})}\BibitemShut {NoStop}%
\bibitem [{\citenamefont {Schulz}\ and\ \citenamefont {Thiemann}(1977)}]{schulz1977crystal}%
  \BibitemOpen
  \bibfield  {author} {\bibinfo {author} {\bibfnamefont {H.}~\bibnamefont {Schulz}}\ and\ \bibinfo {author} {\bibfnamefont {K.}~\bibnamefont {Thiemann}},\ }\bibfield  {title} {\bibinfo {title} {Crystal structure refinement of {A}l{N} and {G}a{N}},\ }\href@noop {} {\bibfield  {journal} {\bibinfo  {journal} {Solid State Commun.}\ }\textbf {\bibinfo {volume} {23}},\ \bibinfo {pages} {815} (\bibinfo {year} {1977})}\BibitemShut {NoStop}%
\bibitem [{\citenamefont {Ruf}\ \emph {et~al.}(2001)\citenamefont {Ruf}, \citenamefont {Serrano}, \citenamefont {Cardona}, \citenamefont {Pavone}, \citenamefont {Pabst}, \citenamefont {Krisch}, \citenamefont {D'astuto}, \citenamefont {Suski}, \citenamefont {Grzegory},\ and\ \citenamefont {Leszczynski}}]{ruf2001phonon}%
  \BibitemOpen
  \bibfield  {author} {\bibinfo {author} {\bibfnamefont {T.}~\bibnamefont {Ruf}}, \bibinfo {author} {\bibfnamefont {J.}~\bibnamefont {Serrano}}, \bibinfo {author} {\bibfnamefont {M.}~\bibnamefont {Cardona}}, \bibinfo {author} {\bibfnamefont {P.}~\bibnamefont {Pavone}}, \bibinfo {author} {\bibfnamefont {M.}~\bibnamefont {Pabst}}, \bibinfo {author} {\bibfnamefont {M.}~\bibnamefont {Krisch}}, \bibinfo {author} {\bibfnamefont {M.}~\bibnamefont {D'astuto}}, \bibinfo {author} {\bibfnamefont {T.}~\bibnamefont {Suski}}, \bibinfo {author} {\bibfnamefont {I.}~\bibnamefont {Grzegory}},\ and\ \bibinfo {author} {\bibfnamefont {M.}~\bibnamefont {Leszczynski}},\ }\bibfield  {title} {\bibinfo {title} {Phonon dispersion curves in wurtzite-structure {G}a{N} determined by inelastic x-ray scattering},\ }\href@noop {} {\bibfield  {journal} {\bibinfo  {journal} {Phys. Rev. Lett.}\ }\textbf {\bibinfo {volume} {86}},\ \bibinfo {pages} {906} (\bibinfo {year} {2001})}\BibitemShut {NoStop}%
\bibitem [{\citenamefont {Tamura}(1983)}]{tamura1983isotope}%
  \BibitemOpen
  \bibfield  {author} {\bibinfo {author} {\bibfnamefont {S.-i.}\ \bibnamefont {Tamura}},\ }\bibfield  {title} {\bibinfo {title} {Isotope scattering of dispersive phonons in {G}e},\ }\href@noop {} {\bibfield  {journal} {\bibinfo  {journal} {Phys. Rev. B}\ }\textbf {\bibinfo {volume} {27}},\ \bibinfo {pages} {858} (\bibinfo {year} {1983})}\BibitemShut {NoStop}%
\bibitem [{\citenamefont {Ponc{\'e}}\ \emph {et~al.}(2016)\citenamefont {Ponc{\'e}}, \citenamefont {Margine}, \citenamefont {Verdi},\ and\ \citenamefont {Giustino}}]{ponce2016epw}%
  \BibitemOpen
  \bibfield  {author} {\bibinfo {author} {\bibfnamefont {S.}~\bibnamefont {Ponc{\'e}}}, \bibinfo {author} {\bibfnamefont {E.~R.}\ \bibnamefont {Margine}}, \bibinfo {author} {\bibfnamefont {C.}~\bibnamefont {Verdi}},\ and\ \bibinfo {author} {\bibfnamefont {F.}~\bibnamefont {Giustino}},\ }\bibfield  {title} {\bibinfo {title} {{EPW}: Electron--phonon coupling, transport and superconducting properties using maximally localized {W}annier functions},\ }\href@noop {} {\bibfield  {journal} {\bibinfo  {journal} {Comput. Phys. Commun.}\ }\textbf {\bibinfo {volume} {209}},\ \bibinfo {pages} {116} (\bibinfo {year} {2016})}\BibitemShut {NoStop}%
\bibitem [{\citenamefont {Giannozzi}\ \emph {et~al.}(2009)\citenamefont {Giannozzi}, \citenamefont {Baroni}, \citenamefont {Bonini}, \citenamefont {Calandra}, \citenamefont {Car}, \citenamefont {Cavazzoni}, \citenamefont {Ceresoli}, \citenamefont {Chiarotti}, \citenamefont {Cococcioni}, \citenamefont {Dabo} \emph {et~al.}}]{giannozzi2009quantum}%
  \BibitemOpen
  \bibfield  {author} {\bibinfo {author} {\bibfnamefont {P.}~\bibnamefont {Giannozzi}}, \bibinfo {author} {\bibfnamefont {S.}~\bibnamefont {Baroni}}, \bibinfo {author} {\bibfnamefont {N.}~\bibnamefont {Bonini}}, \bibinfo {author} {\bibfnamefont {M.}~\bibnamefont {Calandra}}, \bibinfo {author} {\bibfnamefont {R.}~\bibnamefont {Car}}, \bibinfo {author} {\bibfnamefont {C.}~\bibnamefont {Cavazzoni}}, \bibinfo {author} {\bibfnamefont {D.}~\bibnamefont {Ceresoli}}, \bibinfo {author} {\bibfnamefont {G.~L.}\ \bibnamefont {Chiarotti}}, \bibinfo {author} {\bibfnamefont {M.}~\bibnamefont {Cococcioni}}, \bibinfo {author} {\bibfnamefont {I.}~\bibnamefont {Dabo}}, \emph {et~al.},\ }\bibfield  {title} {\bibinfo {title} {{QUANTUM ESPRESSO}: a modular and open-source software project for quantum simulations of materials},\ }\href@noop {} {\bibfield  {journal} {\bibinfo  {journal} {J. Phys.: Condens.Matter}\ }\textbf {\bibinfo {volume} {21}},\ \bibinfo {pages} {395502} (\bibinfo {year} {2009})}\BibitemShut {NoStop}%
\bibitem [{\citenamefont {Perdew}\ \emph {et~al.}(2008)\citenamefont {Perdew}, \citenamefont {Ruzsinszky}, \citenamefont {Csonka}, \citenamefont {Vydrov}, \citenamefont {Scuseria}, \citenamefont {Constantin}, \citenamefont {Zhou},\ and\ \citenamefont {Burke}}]{perdew2008restoring}%
  \BibitemOpen
  \bibfield  {author} {\bibinfo {author} {\bibfnamefont {J.~P.}\ \bibnamefont {Perdew}}, \bibinfo {author} {\bibfnamefont {A.}~\bibnamefont {Ruzsinszky}}, \bibinfo {author} {\bibfnamefont {G.~I.}\ \bibnamefont {Csonka}}, \bibinfo {author} {\bibfnamefont {O.~A.}\ \bibnamefont {Vydrov}}, \bibinfo {author} {\bibfnamefont {G.~E.}\ \bibnamefont {Scuseria}}, \bibinfo {author} {\bibfnamefont {L.~A.}\ \bibnamefont {Constantin}}, \bibinfo {author} {\bibfnamefont {X.}~\bibnamefont {Zhou}},\ and\ \bibinfo {author} {\bibfnamefont {K.}~\bibnamefont {Burke}},\ }\bibfield  {title} {\bibinfo {title} {Restoring the density-gradient expansion for exchange in solids and surfaces},\ }\href@noop {} {\bibfield  {journal} {\bibinfo  {journal} {Phys. Rev. Lett.}\ }\textbf {\bibinfo {volume} {100}},\ \bibinfo {pages} {136406} (\bibinfo {year} {2008})}\BibitemShut {NoStop}%
\bibitem [{\citenamefont {Hamann}(2013)}]{hamann2013optimized}%
  \BibitemOpen
  \bibfield  {author} {\bibinfo {author} {\bibfnamefont {D.}~\bibnamefont {Hamann}},\ }\bibfield  {title} {\bibinfo {title} {Optimized norm-conserving {V}anderbilt pseudopotentials},\ }\href@noop {} {\bibfield  {journal} {\bibinfo  {journal} {Phys. Rev. B}\ }\textbf {\bibinfo {volume} {88}},\ \bibinfo {pages} {085117} (\bibinfo {year} {2013})}\BibitemShut {NoStop}%
\bibitem [{\citenamefont {van Setten}\ \emph {et~al.}(2018)\citenamefont {van Setten}, \citenamefont {Giantomassi}, \citenamefont {Bousquet}, \citenamefont {Verstraete}, \citenamefont {Hamann}, \citenamefont {Gonze},\ and\ \citenamefont {Rignanese}}]{van2018pseudodojo}%
  \BibitemOpen
  \bibfield  {author} {\bibinfo {author} {\bibfnamefont {M.~J.}\ \bibnamefont {van Setten}}, \bibinfo {author} {\bibfnamefont {M.}~\bibnamefont {Giantomassi}}, \bibinfo {author} {\bibfnamefont {E.}~\bibnamefont {Bousquet}}, \bibinfo {author} {\bibfnamefont {M.~J.}\ \bibnamefont {Verstraete}}, \bibinfo {author} {\bibfnamefont {D.~R.}\ \bibnamefont {Hamann}}, \bibinfo {author} {\bibfnamefont {X.}~\bibnamefont {Gonze}},\ and\ \bibinfo {author} {\bibfnamefont {G.-M.}\ \bibnamefont {Rignanese}},\ }\bibfield  {title} {\bibinfo {title} {The {P}seudo{D}ojo: {T}raining and grading a 85 element optimized norm-conserving pseudopotential table},\ }\href@noop {} {\bibfield  {journal} {\bibinfo  {journal} {Comput. Phys. Commun.}\ }\textbf {\bibinfo {volume} {226}},\ \bibinfo {pages} {39} (\bibinfo {year} {2018})}\BibitemShut {NoStop}%
\bibitem [{\citenamefont {Ponc{\'e}}\ \emph {et~al.}(2019{\natexlab{a}})\citenamefont {Ponc{\'e}}, \citenamefont {Jena},\ and\ \citenamefont {Giustino}}]{ponce2019route}%
  \BibitemOpen
  \bibfield  {author} {\bibinfo {author} {\bibfnamefont {S.}~\bibnamefont {Ponc{\'e}}}, \bibinfo {author} {\bibfnamefont {D.}~\bibnamefont {Jena}},\ and\ \bibinfo {author} {\bibfnamefont {F.}~\bibnamefont {Giustino}},\ }\bibfield  {title} {\bibinfo {title} {Route to high hole mobility in {G}a{N} via reversal of crystal-field splitting},\ }\href@noop {} {\bibfield  {journal} {\bibinfo  {journal} {Phys. Rev. Lett.}\ }\textbf {\bibinfo {volume} {123}},\ \bibinfo {pages} {096602} (\bibinfo {year} {2019}{\natexlab{a}})}\BibitemShut {NoStop}%
\bibitem [{\citenamefont {Marzari}\ \emph {et~al.}(2012)\citenamefont {Marzari}, \citenamefont {Mostofi}, \citenamefont {Yates}, \citenamefont {Souza},\ and\ \citenamefont {Vanderbilt}}]{marzari2012maximally}%
  \BibitemOpen
  \bibfield  {author} {\bibinfo {author} {\bibfnamefont {N.}~\bibnamefont {Marzari}}, \bibinfo {author} {\bibfnamefont {A.~A.}\ \bibnamefont {Mostofi}}, \bibinfo {author} {\bibfnamefont {J.~R.}\ \bibnamefont {Yates}}, \bibinfo {author} {\bibfnamefont {I.}~\bibnamefont {Souza}},\ and\ \bibinfo {author} {\bibfnamefont {D.}~\bibnamefont {Vanderbilt}},\ }\bibfield  {title} {\bibinfo {title} {Maximally localized {W}annier functions: {T}heory and applications},\ }\href@noop {} {\bibfield  {journal} {\bibinfo  {journal} {Rev. Mod. Phys.}\ }\textbf {\bibinfo {volume} {84}},\ \bibinfo {pages} {1419} (\bibinfo {year} {2012})}\BibitemShut {NoStop}%
\bibitem [{\citenamefont {Paulatto}\ \emph {et~al.}(2013)\citenamefont {Paulatto}, \citenamefont {Mauri},\ and\ \citenamefont {Lazzeri}}]{paulatto2013anharmonic}%
  \BibitemOpen
  \bibfield  {author} {\bibinfo {author} {\bibfnamefont {L.}~\bibnamefont {Paulatto}}, \bibinfo {author} {\bibfnamefont {F.}~\bibnamefont {Mauri}},\ and\ \bibinfo {author} {\bibfnamefont {M.}~\bibnamefont {Lazzeri}},\ }\bibfield  {title} {\bibinfo {title} {Anharmonic properties from a generalized third-order ab initio approach: Theory and applications to graphite and graphene},\ }\href@noop {} {\bibfield  {journal} {\bibinfo  {journal} {Phys. Rev. B}\ }\textbf {\bibinfo {volume} {87}},\ \bibinfo {pages} {214303} (\bibinfo {year} {2013})}\BibitemShut {NoStop}%
\bibitem [{\citenamefont {Li}\ \emph {et~al.}(2022)\citenamefont {Li}, \citenamefont {Tong}, \citenamefont {Shao}, \citenamefont {Bao}, \citenamefont {Frauenheim},\ and\ \citenamefont {Liu}}]{li2022anomalously}%
  \BibitemOpen
  \bibfield  {author} {\bibinfo {author} {\bibfnamefont {S.}~\bibnamefont {Li}}, \bibinfo {author} {\bibfnamefont {Z.}~\bibnamefont {Tong}}, \bibinfo {author} {\bibfnamefont {C.}~\bibnamefont {Shao}}, \bibinfo {author} {\bibfnamefont {H.}~\bibnamefont {Bao}}, \bibinfo {author} {\bibfnamefont {T.}~\bibnamefont {Frauenheim}},\ and\ \bibinfo {author} {\bibfnamefont {X.}~\bibnamefont {Liu}},\ }\bibfield  {title} {\bibinfo {title} {Anomalously isotropic electron transport and weak electron--phonon interactions in hexagonal noble metals},\ }\href@noop {} {\bibfield  {journal} {\bibinfo  {journal} {J. Phys. Chem. Lett.}\ }\textbf {\bibinfo {volume} {13}},\ \bibinfo {pages} {4289} (\bibinfo {year} {2022})}\BibitemShut {NoStop}%
\bibitem [{\citenamefont {Li}\ \emph {et~al.}(2020)\citenamefont {Li}, \citenamefont {Wang}, \citenamefont {Hu}, \citenamefont {Gu}, \citenamefont {Tong},\ and\ \citenamefont {Bao}}]{li2020anomalous}%
  \BibitemOpen
  \bibfield  {author} {\bibinfo {author} {\bibfnamefont {S.}~\bibnamefont {Li}}, \bibinfo {author} {\bibfnamefont {A.}~\bibnamefont {Wang}}, \bibinfo {author} {\bibfnamefont {Y.}~\bibnamefont {Hu}}, \bibinfo {author} {\bibfnamefont {X.}~\bibnamefont {Gu}}, \bibinfo {author} {\bibfnamefont {Z.}~\bibnamefont {Tong}},\ and\ \bibinfo {author} {\bibfnamefont {H.}~\bibnamefont {Bao}},\ }\bibfield  {title} {\bibinfo {title} {Anomalous thermal transport in metallic transition-metal nitrides originated from strong electron--phonon interactions},\ }\href@noop {} {\bibfield  {journal} {\bibinfo  {journal} {Mater. Today Phys.}\ }\textbf {\bibinfo {volume} {15}},\ \bibinfo {pages} {100256} (\bibinfo {year} {2020})}\BibitemShut {NoStop}%
\bibitem [{\citenamefont {Sun}\ \emph {et~al.}(2023)\citenamefont {Sun}, \citenamefont {Chen}, \citenamefont {Li},\ and\ \citenamefont {Liu}}]{sun2023light}%
  \BibitemOpen
  \bibfield  {author} {\bibinfo {author} {\bibfnamefont {J.}~\bibnamefont {Sun}}, \bibinfo {author} {\bibfnamefont {G.}~\bibnamefont {Chen}}, \bibinfo {author} {\bibfnamefont {S.}~\bibnamefont {Li}},\ and\ \bibinfo {author} {\bibfnamefont {X.}~\bibnamefont {Liu}},\ }\bibfield  {title} {\bibinfo {title} {Light atomic mass induces low lattice thermal conductivity in {J}anus transition-metal dichalcogenides {MSS}e ({M= M}o, {W})},\ }\href@noop {} {\bibfield  {journal} {\bibinfo  {journal} {J. Phys. Chem. C}\ }\textbf {\bibinfo {volume} {127}},\ \bibinfo {pages} {17567} (\bibinfo {year} {2023})}\BibitemShut {NoStop}%
\bibitem [{\citenamefont {Li}\ \emph {et~al.}(2019)\citenamefont {Li}, \citenamefont {Tong},\ and\ \citenamefont {Bao}}]{li2019resolving}%
  \BibitemOpen
  \bibfield  {author} {\bibinfo {author} {\bibfnamefont {S.}~\bibnamefont {Li}}, \bibinfo {author} {\bibfnamefont {Z.}~\bibnamefont {Tong}},\ and\ \bibinfo {author} {\bibfnamefont {H.}~\bibnamefont {Bao}},\ }\bibfield  {title} {\bibinfo {title} {Resolving different scattering effects on the thermal and electrical transport in doped {S}n{S}e},\ }\href@noop {} {\bibfield  {journal} {\bibinfo  {journal} {J. Appl. Phys.}\ }\textbf {\bibinfo {volume} {126}},\ \bibinfo {pages} {025111} (\bibinfo {year} {2019})}\BibitemShut {NoStop}%
\bibitem [{\citenamefont {Dingle}\ \emph {et~al.}(1971)\citenamefont {Dingle}, \citenamefont {Sell}, \citenamefont {Stokowski},\ and\ \citenamefont {Ilegems}}]{dingle1971absorption}%
  \BibitemOpen
  \bibfield  {author} {\bibinfo {author} {\bibfnamefont {R.}~\bibnamefont {Dingle}}, \bibinfo {author} {\bibfnamefont {D.}~\bibnamefont {Sell}}, \bibinfo {author} {\bibfnamefont {S.}~\bibnamefont {Stokowski}},\ and\ \bibinfo {author} {\bibfnamefont {M.}~\bibnamefont {Ilegems}},\ }\bibfield  {title} {\bibinfo {title} {Absorption, reflectance, and luminescence of {G}a{N} epitaxial layers},\ }\href@noop {} {\bibfield  {journal} {\bibinfo  {journal} {Phys. Rev. B}\ }\textbf {\bibinfo {volume} {4}},\ \bibinfo {pages} {1211} (\bibinfo {year} {1971})}\BibitemShut {NoStop}%
\bibitem [{\citenamefont {Monemar}(1974)}]{monemar1974fundamental}%
  \BibitemOpen
  \bibfield  {author} {\bibinfo {author} {\bibfnamefont {B.}~\bibnamefont {Monemar}},\ }\bibfield  {title} {\bibinfo {title} {Fundamental energy gap of {G}a{N} from photoluminescence excitation spectra},\ }\href@noop {} {\bibfield  {journal} {\bibinfo  {journal} {Phys. Rev. B}\ }\textbf {\bibinfo {volume} {10}},\ \bibinfo {pages} {676} (\bibinfo {year} {1974})}\BibitemShut {NoStop}%
\bibitem [{\citenamefont {Chan}\ and\ \citenamefont {Ceder}(2010)}]{chan2010efficient}%
  \BibitemOpen
  \bibfield  {author} {\bibinfo {author} {\bibfnamefont {M.~K.}\ \bibnamefont {Chan}}\ and\ \bibinfo {author} {\bibfnamefont {G.}~\bibnamefont {Ceder}},\ }\bibfield  {title} {\bibinfo {title} {Efficient band gap prediction for solids},\ }\href@noop {} {\bibfield  {journal} {\bibinfo  {journal} {Phys. Rev. Lett.}\ }\textbf {\bibinfo {volume} {105}},\ \bibinfo {pages} {196403} (\bibinfo {year} {2010})}\BibitemShut {NoStop}%
\bibitem [{\citenamefont {Rodina}\ \emph {et~al.}(2001)\citenamefont {Rodina}, \citenamefont {Dietrich}, \citenamefont {G{\"o}ldner}, \citenamefont {Eckey}, \citenamefont {Hoffmann}, \citenamefont {Efros}, \citenamefont {Rosen},\ and\ \citenamefont {Meyer}}]{rodina2001free}%
  \BibitemOpen
  \bibfield  {author} {\bibinfo {author} {\bibfnamefont {A.}~\bibnamefont {Rodina}}, \bibinfo {author} {\bibfnamefont {M.}~\bibnamefont {Dietrich}}, \bibinfo {author} {\bibfnamefont {A.}~\bibnamefont {G{\"o}ldner}}, \bibinfo {author} {\bibfnamefont {L.}~\bibnamefont {Eckey}}, \bibinfo {author} {\bibfnamefont {A.}~\bibnamefont {Hoffmann}}, \bibinfo {author} {\bibfnamefont {A.~L.}\ \bibnamefont {Efros}}, \bibinfo {author} {\bibfnamefont {M.}~\bibnamefont {Rosen}},\ and\ \bibinfo {author} {\bibfnamefont {B.}~\bibnamefont {Meyer}},\ }\bibfield  {title} {\bibinfo {title} {Free excitons in wurtzite {G}a{N}},\ }\href@noop {} {\bibfield  {journal} {\bibinfo  {journal} {Phys. Rev. B}\ }\textbf {\bibinfo {volume} {64}},\ \bibinfo {pages} {115204} (\bibinfo {year} {2001})}\BibitemShut {NoStop}%
\bibitem [{\citenamefont {Je{\.z}owski}\ \emph {et~al.}(2003)\citenamefont {Je{\.z}owski}, \citenamefont {Stachowiak}, \citenamefont {Plackowski}, \citenamefont {Suski}, \citenamefont {Krukowski}, \citenamefont {Bo{\'c}kowski}, \citenamefont {Grzegory}, \citenamefont {Danilchenko},\ and\ \citenamefont {Paszkiewicz}}]{jezowski2003thermal}%
  \BibitemOpen
  \bibfield  {author} {\bibinfo {author} {\bibfnamefont {A.}~\bibnamefont {Je{\.z}owski}}, \bibinfo {author} {\bibfnamefont {P.}~\bibnamefont {Stachowiak}}, \bibinfo {author} {\bibfnamefont {T.}~\bibnamefont {Plackowski}}, \bibinfo {author} {\bibfnamefont {T.}~\bibnamefont {Suski}}, \bibinfo {author} {\bibfnamefont {S.}~\bibnamefont {Krukowski}}, \bibinfo {author} {\bibfnamefont {M.}~\bibnamefont {Bo{\'c}kowski}}, \bibinfo {author} {\bibfnamefont {I.}~\bibnamefont {Grzegory}}, \bibinfo {author} {\bibfnamefont {B.}~\bibnamefont {Danilchenko}},\ and\ \bibinfo {author} {\bibfnamefont {T.}~\bibnamefont {Paszkiewicz}},\ }\bibfield  {title} {\bibinfo {title} {Thermal conductivity of {G}a{N} crystals grown by high pressure method},\ }\href@noop {} {\bibfield  {journal} {\bibinfo  {journal} {Phys. Status Solidi B}\ }\textbf {\bibinfo {volume} {240}},\ \bibinfo {pages} {447} (\bibinfo {year} {2003})}\BibitemShut {NoStop}%
\bibitem [{\citenamefont {Slack}\ \emph {et~al.}(2002)\citenamefont {Slack}, \citenamefont {Schowalter}, \citenamefont {Morelli},\ and\ \citenamefont {Freitas~Jr}}]{slack2002some}%
  \BibitemOpen
  \bibfield  {author} {\bibinfo {author} {\bibfnamefont {G.~A.}\ \bibnamefont {Slack}}, \bibinfo {author} {\bibfnamefont {L.~J.}\ \bibnamefont {Schowalter}}, \bibinfo {author} {\bibfnamefont {D.}~\bibnamefont {Morelli}},\ and\ \bibinfo {author} {\bibfnamefont {J.~A.}\ \bibnamefont {Freitas~Jr}},\ }\bibfield  {title} {\bibinfo {title} {Some effects of oxygen impurities on {A}l{N} and {G}a{N}},\ }\href@noop {} {\bibfield  {journal} {\bibinfo  {journal} {J. Cryst. Growth}\ }\textbf {\bibinfo {volume} {246}},\ \bibinfo {pages} {287} (\bibinfo {year} {2002})}\BibitemShut {NoStop}%
\bibitem [{\citenamefont {Pang}\ \emph {et~al.}(2024)\citenamefont {Pang}, \citenamefont {Meng}, \citenamefont {Chen}, \citenamefont {Katre}, \citenamefont {Carrete}, \citenamefont {Dongre}, \citenamefont {Madsen}, \citenamefont {Mingo},\ and\ \citenamefont {Li}}]{pang2024thermal}%
  \BibitemOpen
  \bibfield  {author} {\bibinfo {author} {\bibfnamefont {G.}~\bibnamefont {Pang}}, \bibinfo {author} {\bibfnamefont {F.}~\bibnamefont {Meng}}, \bibinfo {author} {\bibfnamefont {Y.}~\bibnamefont {Chen}}, \bibinfo {author} {\bibfnamefont {A.}~\bibnamefont {Katre}}, \bibinfo {author} {\bibfnamefont {J.}~\bibnamefont {Carrete}}, \bibinfo {author} {\bibfnamefont {B.}~\bibnamefont {Dongre}}, \bibinfo {author} {\bibfnamefont {G.~K.}\ \bibnamefont {Madsen}}, \bibinfo {author} {\bibfnamefont {N.}~\bibnamefont {Mingo}},\ and\ \bibinfo {author} {\bibfnamefont {W.}~\bibnamefont {Li}},\ }\bibfield  {title} {\bibinfo {title} {Thermal conductivity reduction in highly-doped cubic {S}i{C} by phonon-defect and phonon-electron scattering},\ }\href@noop {} {\bibfield  {journal} {\bibinfo  {journal} {Mater. Today Phys.}\ }\textbf {\bibinfo {volume} {41}},\ \bibinfo {pages} {101346} (\bibinfo {year} {2024})}\BibitemShut {NoStop}%
\bibitem [{\citenamefont {Wu}\ \emph {et~al.}(2016)\citenamefont {Wu}, \citenamefont {Lee}, \citenamefont {Varshney}, \citenamefont {Wohlwend}, \citenamefont {Roy},\ and\ \citenamefont {Luo}}]{wu2016thermal}%
  \BibitemOpen
  \bibfield  {author} {\bibinfo {author} {\bibfnamefont {X.}~\bibnamefont {Wu}}, \bibinfo {author} {\bibfnamefont {J.}~\bibnamefont {Lee}}, \bibinfo {author} {\bibfnamefont {V.}~\bibnamefont {Varshney}}, \bibinfo {author} {\bibfnamefont {J.~L.}\ \bibnamefont {Wohlwend}}, \bibinfo {author} {\bibfnamefont {A.~K.}\ \bibnamefont {Roy}},\ and\ \bibinfo {author} {\bibfnamefont {T.}~\bibnamefont {Luo}},\ }\bibfield  {title} {\bibinfo {title} {Thermal conductivity of wurtzite zinc-oxide from first-principles lattice dynamics--a comparative study with gallium nitride},\ }\href@noop {} {\bibfield  {journal} {\bibinfo  {journal} {Sci. Rep.}\ }\textbf {\bibinfo {volume} {6}},\ \bibinfo {pages} {22504} (\bibinfo {year} {2016})}\BibitemShut {NoStop}%
\bibitem [{\citenamefont {Garg}\ \emph {et~al.}(2018)\citenamefont {Garg}, \citenamefont {Luo},\ and\ \citenamefont {Chen}}]{garg2018spectral}%
  \BibitemOpen
  \bibfield  {author} {\bibinfo {author} {\bibfnamefont {J.}~\bibnamefont {Garg}}, \bibinfo {author} {\bibfnamefont {T.}~\bibnamefont {Luo}},\ and\ \bibinfo {author} {\bibfnamefont {G.}~\bibnamefont {Chen}},\ }\bibfield  {title} {\bibinfo {title} {Spectral concentration of thermal conductivity in {G}a{N}—{A} first-principles study},\ }\href@noop {} {\bibfield  {journal} {\bibinfo  {journal} {Appl. Phys. Lett.}\ }\textbf {\bibinfo {volume} {112}},\ \bibinfo {pages} {252101} (\bibinfo {year} {2018})}\BibitemShut {NoStop}%
\bibitem [{\citenamefont {Sheih}\ \emph {et~al.}(1995)\citenamefont {Sheih}, \citenamefont {Tsen}, \citenamefont {Ferry}, \citenamefont {Botchkarev}, \citenamefont {Sverdlov}, \citenamefont {Salvador},\ and\ \citenamefont {Morkoc}}]{sheih1995electron}%
  \BibitemOpen
  \bibfield  {author} {\bibinfo {author} {\bibfnamefont {S.}~\bibnamefont {Sheih}}, \bibinfo {author} {\bibfnamefont {K.-T.}\ \bibnamefont {Tsen}}, \bibinfo {author} {\bibfnamefont {D.}~\bibnamefont {Ferry}}, \bibinfo {author} {\bibfnamefont {A.}~\bibnamefont {Botchkarev}}, \bibinfo {author} {\bibfnamefont {B.}~\bibnamefont {Sverdlov}}, \bibinfo {author} {\bibfnamefont {A.}~\bibnamefont {Salvador}},\ and\ \bibinfo {author} {\bibfnamefont {H.}~\bibnamefont {Morkoc}},\ }\bibfield  {title} {\bibinfo {title} {Electron--phonon interactions in the wide band-gap semiconductor {G}a{N}},\ }\href@noop {} {\bibfield  {journal} {\bibinfo  {journal} {Appl. Phys. Lett.}\ }\textbf {\bibinfo {volume} {67}},\ \bibinfo {pages} {1757} (\bibinfo {year} {1995})}\BibitemShut {NoStop}%
\bibitem [{\citenamefont {Yoodee}\ \emph {et~al.}(1984)\citenamefont {Yoodee}, \citenamefont {Woolley},\ and\ \citenamefont {Sa-Yakanit}}]{yoodee1984effects}%
  \BibitemOpen
  \bibfield  {author} {\bibinfo {author} {\bibfnamefont {K.}~\bibnamefont {Yoodee}}, \bibinfo {author} {\bibfnamefont {J.~C.}\ \bibnamefont {Woolley}},\ and\ \bibinfo {author} {\bibfnamefont {V.}~\bibnamefont {Sa-Yakanit}},\ }\bibfield  {title} {\bibinfo {title} {Effects of p-d hybridization on the valence band of {I-III-VI}$_{2}$ chalcopyrite semiconductors},\ }\href@noop {} {\bibfield  {journal} {\bibinfo  {journal} {Phys. Rev. B}\ }\textbf {\bibinfo {volume} {30}},\ \bibinfo {pages} {5904} (\bibinfo {year} {1984})}\BibitemShut {NoStop}%
\bibitem [{\citenamefont {Reynolds}\ \emph {et~al.}(1999)\citenamefont {Reynolds}, \citenamefont {Look}, \citenamefont {Jogai}, \citenamefont {Litton}, \citenamefont {Cantwell},\ and\ \citenamefont {Harsch}}]{reynolds1999valence}%
  \BibitemOpen
  \bibfield  {author} {\bibinfo {author} {\bibfnamefont {D.}~\bibnamefont {Reynolds}}, \bibinfo {author} {\bibfnamefont {D.~C.}\ \bibnamefont {Look}}, \bibinfo {author} {\bibfnamefont {B.}~\bibnamefont {Jogai}}, \bibinfo {author} {\bibfnamefont {C.}~\bibnamefont {Litton}}, \bibinfo {author} {\bibfnamefont {G.}~\bibnamefont {Cantwell}},\ and\ \bibinfo {author} {\bibfnamefont {W.}~\bibnamefont {Harsch}},\ }\bibfield  {title} {\bibinfo {title} {Valence-band ordering in {Z}n{O}},\ }\href@noop {} {\bibfield  {journal} {\bibinfo  {journal} {Phys. Rev. B}\ }\textbf {\bibinfo {volume} {60}},\ \bibinfo {pages} {2340} (\bibinfo {year} {1999})}\BibitemShut {NoStop}%
\bibitem [{\citenamefont {Walsh}\ \emph {et~al.}(2008)\citenamefont {Walsh}, \citenamefont {Da~Silva},\ and\ \citenamefont {Wei}}]{walsh2008origins}%
  \BibitemOpen
  \bibfield  {author} {\bibinfo {author} {\bibfnamefont {A.}~\bibnamefont {Walsh}}, \bibinfo {author} {\bibfnamefont {J.~L.}\ \bibnamefont {Da~Silva}},\ and\ \bibinfo {author} {\bibfnamefont {S.-H.}\ \bibnamefont {Wei}},\ }\bibfield  {title} {\bibinfo {title} {Origins of band-gap renormalization in degenerately doped semiconductors},\ }\href@noop {} {\bibfield  {journal} {\bibinfo  {journal} {Phys. Rev. B}\ }\textbf {\bibinfo {volume} {78}},\ \bibinfo {pages} {075211} (\bibinfo {year} {2008})}\BibitemShut {NoStop}%
\bibitem [{\citenamefont {Wang}\ \emph {et~al.}(2016)\citenamefont {Wang}, \citenamefont {Lu},\ and\ \citenamefont {Ruan}}]{wang2016first}%
  \BibitemOpen
  \bibfield  {author} {\bibinfo {author} {\bibfnamefont {Y.}~\bibnamefont {Wang}}, \bibinfo {author} {\bibfnamefont {Z.}~\bibnamefont {Lu}},\ and\ \bibinfo {author} {\bibfnamefont {X.}~\bibnamefont {Ruan}},\ }\bibfield  {title} {\bibinfo {title} {First principles calculation of lattice thermal conductivity of metals considering phonon-phonon and phonon-electron scattering},\ }\href@noop {} {\bibfield  {journal} {\bibinfo  {journal} {J. Appl. Phys.}\ }\textbf {\bibinfo {volume} {119}},\ \bibinfo {pages} {225109} (\bibinfo {year} {2016})}\BibitemShut {NoStop}%
\bibitem [{\citenamefont {Li}\ \emph {et~al.}(2021)\citenamefont {Li}, \citenamefont {Zhang},\ and\ \citenamefont {Bao}}]{li2021thermal}%
  \BibitemOpen
  \bibfield  {author} {\bibinfo {author} {\bibfnamefont {S.}~\bibnamefont {Li}}, \bibinfo {author} {\bibfnamefont {X.}~\bibnamefont {Zhang}},\ and\ \bibinfo {author} {\bibfnamefont {H.}~\bibnamefont {Bao}},\ }\bibfield  {title} {\bibinfo {title} {Thermal transport by electrons and phonons in {P}d{T}e$_{2}$: an ab initio study},\ }\href@noop {} {\bibfield  {journal} {\bibinfo  {journal} {Phys. Chem. Chem. Phys.}\ }\textbf {\bibinfo {volume} {23}},\ \bibinfo {pages} {5956} (\bibinfo {year} {2021})}\BibitemShut {NoStop}%
\bibitem [{\citenamefont {Li}\ \emph {et~al.}(2018)\citenamefont {Li}, \citenamefont {Ravichandran}, \citenamefont {Lindsay},\ and\ \citenamefont {Broido}}]{li2018fermi}%
  \BibitemOpen
  \bibfield  {author} {\bibinfo {author} {\bibfnamefont {C.}~\bibnamefont {Li}}, \bibinfo {author} {\bibfnamefont {N.~K.}\ \bibnamefont {Ravichandran}}, \bibinfo {author} {\bibfnamefont {L.}~\bibnamefont {Lindsay}},\ and\ \bibinfo {author} {\bibfnamefont {D.}~\bibnamefont {Broido}},\ }\bibfield  {title} {\bibinfo {title} {Fermi surface nesting and phonon frequency gap drive anomalous thermal transport},\ }\href@noop {} {\bibfield  {journal} {\bibinfo  {journal} {Phys. Rev. Lett.}\ }\textbf {\bibinfo {volume} {121}},\ \bibinfo {pages} {175901} (\bibinfo {year} {2018})}\BibitemShut {NoStop}%
\bibitem [{\citenamefont {Kasinathan}\ \emph {et~al.}(2006)\citenamefont {Kasinathan}, \citenamefont {Kune{\v{s}}}, \citenamefont {Lazicki}, \citenamefont {Rosner}, \citenamefont {Yoo}, \citenamefont {Scalettar},\ and\ \citenamefont {Pickett}}]{kasinathan2006superconductivity}%
  \BibitemOpen
  \bibfield  {author} {\bibinfo {author} {\bibfnamefont {D.}~\bibnamefont {Kasinathan}}, \bibinfo {author} {\bibfnamefont {J.}~\bibnamefont {Kune{\v{s}}}}, \bibinfo {author} {\bibfnamefont {A.}~\bibnamefont {Lazicki}}, \bibinfo {author} {\bibfnamefont {H.}~\bibnamefont {Rosner}}, \bibinfo {author} {\bibfnamefont {C.}~\bibnamefont {Yoo}}, \bibinfo {author} {\bibfnamefont {R.}~\bibnamefont {Scalettar}},\ and\ \bibinfo {author} {\bibfnamefont {W.}~\bibnamefont {Pickett}},\ }\bibfield  {title} {\bibinfo {title} {Superconductivity and lattice instability in compressed lithium from {F}ermi surface hot spots},\ }\href@noop {} {\bibfield  {journal} {\bibinfo  {journal} {Phys. Rev. Lett.}\ }\textbf {\bibinfo {volume} {96}},\ \bibinfo {pages} {047004} (\bibinfo {year} {2006})}\BibitemShut {NoStop}%
\bibitem [{\citenamefont {Curtarolo}\ \emph {et~al.}(2012)\citenamefont {Curtarolo}, \citenamefont {Setyawan}, \citenamefont {Hart}, \citenamefont {Jahnatek}, \citenamefont {Chepulskii}, \citenamefont {Taylor}, \citenamefont {Wang}, \citenamefont {Xue}, \citenamefont {Yang}, \citenamefont {Levy} \emph {et~al.}}]{curtarolo2012aflow}%
  \BibitemOpen
  \bibfield  {author} {\bibinfo {author} {\bibfnamefont {S.}~\bibnamefont {Curtarolo}}, \bibinfo {author} {\bibfnamefont {W.}~\bibnamefont {Setyawan}}, \bibinfo {author} {\bibfnamefont {G.~L.}\ \bibnamefont {Hart}}, \bibinfo {author} {\bibfnamefont {M.}~\bibnamefont {Jahnatek}}, \bibinfo {author} {\bibfnamefont {R.~V.}\ \bibnamefont {Chepulskii}}, \bibinfo {author} {\bibfnamefont {R.~H.}\ \bibnamefont {Taylor}}, \bibinfo {author} {\bibfnamefont {S.}~\bibnamefont {Wang}}, \bibinfo {author} {\bibfnamefont {J.}~\bibnamefont {Xue}}, \bibinfo {author} {\bibfnamefont {K.}~\bibnamefont {Yang}}, \bibinfo {author} {\bibfnamefont {O.}~\bibnamefont {Levy}}, \emph {et~al.},\ }\bibfield  {title} {\bibinfo {title} {{AFLOW}: {A}n automatic framework for high-throughput materials discovery},\ }\href@noop {} {\bibfield  {journal} {\bibinfo  {journal} {Comput. Mater. Sci.}\ }\textbf {\bibinfo {volume} {58}},\ \bibinfo {pages} {218} (\bibinfo {year} {2012})}\BibitemShut {NoStop}%
\bibitem [{\citenamefont {Ponc{\'e}}\ \emph {et~al.}(2019{\natexlab{b}})\citenamefont {Ponc{\'e}}, \citenamefont {Jena},\ and\ \citenamefont {Giustino}}]{ponce2019hole}%
  \BibitemOpen
  \bibfield  {author} {\bibinfo {author} {\bibfnamefont {S.}~\bibnamefont {Ponc{\'e}}}, \bibinfo {author} {\bibfnamefont {D.}~\bibnamefont {Jena}},\ and\ \bibinfo {author} {\bibfnamefont {F.}~\bibnamefont {Giustino}},\ }\bibfield  {title} {\bibinfo {title} {Hole mobility of strained {G}a{N} from first principles},\ }\href@noop {} {\bibfield  {journal} {\bibinfo  {journal} {Phys. Rev. B}\ }\textbf {\bibinfo {volume} {100}},\ \bibinfo {pages} {085204} (\bibinfo {year} {2019}{\natexlab{b}})}\BibitemShut {NoStop}%
\end{thebibliography}%
\end{document}


\preprint{APS/123-QED}
\renewcommand{\thetable}{S\arabic{table}}
\renewcommand{\thefigure}{S\arabic{figure}}
\renewcommand{\theequation}{S\arabic{equation}}
\title{Weak effects of electron-phonon interactions on the lattice thermal conductivity of wurtzite GaN with high electron concentrations\\[6pt]
Supplemental Material}


\author{Jianshi Sun}
\affiliation{Institute of Micro/Nano Electromechanical System and Integrated Circuit, College of Mechanical Engineering, Donghua University, Shanghai 201620, China
}%

\author{Shouhang Li}
\email{shouhang.li@dhu.edu.cn}
\affiliation{Institute of Micro/Nano Electromechanical System and Integrated Circuit, College of Mechanical Engineering, Donghua University, Shanghai 201620, China
}%

\author{Zhen Tong}
\affiliation{
School of Advanced Energy, Sun Yat-Sen University, Shenzhen 518107, China
}

\author{Cheng Shao}
\affiliation{
 Thermal Science Research Center, Shandong Institute of Advanced Technology, Jinan, Shandong 250103, China
}

\author{Xiangchuan Chen}
\affiliation{%
Institute of Micro/Nano Electromechanical System and Integrated Circuit, College of Mechanical Engineering, Donghua University, Shanghai 201620, China
}

\author{Qianqian Liu}
\affiliation{%
Institute of Micro/Nano Electromechanical System and Integrated Circuit, College of Mechanical Engineering, Donghua University, Shanghai 201620, China
}

\author{Yucheng Xiong}
\affiliation{%
Institute of Micro/Nano Electromechanical System and Integrated Circuit, College of Mechanical Engineering, Donghua University, Shanghai 201620, China
}

\author{Meng An}
\affiliation{
 Department of Mechanical Engineering, The University of Tokyo, 7-3-1 Hongo, Bunkyo, Tokyo, 113-8656, Japan
}

\author{Xiangjun Liu}
\email{xjliu@dhu.edu.cn}
\affiliation{%
 Institute of Micro/Nano Electromechanical System and Integrated Circuit, College of Mechanical Engineering, Donghua University, Shanghai 201620, China
}%


\date{\today}
\maketitle

\section{Computational Details}
The lattice vectors and atomic positions are fully relaxed based on the Broyden-Fretcher-Goldfarb-Shanno (BFGS) optimization \cite{fletcher1970new,broyden1970convergence,goldfarb1970family,shanno1970conditioning} and the convergence thresholds for energy and force are set to $10^{-8}$ Ry and $10^{-8}$ Ry/bohr, respectively. The Brillouin zone is sampled using a $12 \times 12 \times 8$ Monkhorst-Pack \textbf{k}-point mesh. It should be noted that the lattice parameters are marginally underestimated using the local-density approximation (LDA) form of the exchange-correlation functional in Ref. \cite{yang2019stronger}. In contrast, the lattice parameters (\textit{a} = 3.184 Å and \textit{c} = 5.187 Å) calculated from the PBE form of the exchange-correlation functional in this work agree well with the experimental data (\textit{a} = 3.190 Å, \textit{c} = 5.189Å)\cite{schulz1977crystal}. We also calculate the lattice thermal conductivity of GaN using the lattice parameters (\textit{a} = 3.160 Å and \textit{c} = 5.150 Å) in Ref. \cite{yang2019stronger} and relax the atomic positions. Interestingly, the lattice thermal conductivity of GaN decreases from 264 to 232 W/mK. It indicates that the lattice thermal conductivity of GaN is quite sensitive to the lattice parameters. Since our lattice parameters are optimized based on the BFGS optimization method, we do not artificially adjust them to make the lattice thermal conductivity match the existing experimental results better. For the harmonic force constant calculations, the electronic wavevector grid is set to be $12 \times 12 \times 8$. The \textbf{q}-point mesh is set to be $6 \times 6 \times 4$ and the energy threshold is $10^{-17}$ Ry to guarantee the convergence. Furthermore, the dielectric constant and Born effective charge are calculated to account for the long-range electrostatic interactions. For the cubic force constant calculations, the \textbf{q}-point mesh is set to be $3 \times 3 \times 2$. For the calculations of phonon-electron scattering, maximally-localized Wannier functions (MLWFs)\cite{marzari2012maximally} are used for the interpolation in the first Brillouin zone. We adopt the selected columns of the density matrix method (SCDM) to automatically generate MLWFs\cite{damle2015compressed,damle2018disentanglement}. The electron-phonon coupling matrix elements are first calculated under coarse \textbf{k}/\textbf{q} meshes ($12 \times 12 \times 8$/$6 \times 6 \times 4$) and then interpolated to dense meshes ($36 \times 36 \times 24$/$15 \times 15 \times 10$) with Wannier interpolation technique\cite{marzari2012maximally}. $\kappa_{\text {lat}}$ is converged with respect to a \textbf{q}-mesh of $15 \times 15 \times 10$.

The optimized lattice parameter for Si is \textit{a} = 5.392 Å in good agreement with experimental values\cite{straumanis1952lattice}. For the harmonic force constant, the \textbf{q}-point mesh is set to be $6 \times 6 \times 6$ and the energy threshold for phonon calculation is set to be $10^{-17}$ Ry. For the cubic force constant, the \textbf{q}-point mesh is set to be $3 \times 3 \times 3$. For the phonon-electron scattering, the electron-phonon coupling matrix elements are first calculated on coarse meshes of $12 \times 12 \times 12$ \textbf{k}-point mesh and $6 \times 6 \times 6$ \textbf{q}-point mesh and are then interpolated to dense meshes of $60 \times 60 \times 60$ \textbf{k}-point mesh and $40 \times 40 \times 40$ \textbf{q}-point mesh, which is large enough to ensure the convergence. The \textbf{q}-point mesh is set to be $40 \times 40 \times 40$ when solving the phonon Boltzmann transport equation.

\section{The weighted phase space of three-phonon and four-phonon scattering}
\begin{figure}[H]
    \centering
    \includegraphics[width=0.52\columnwidth]{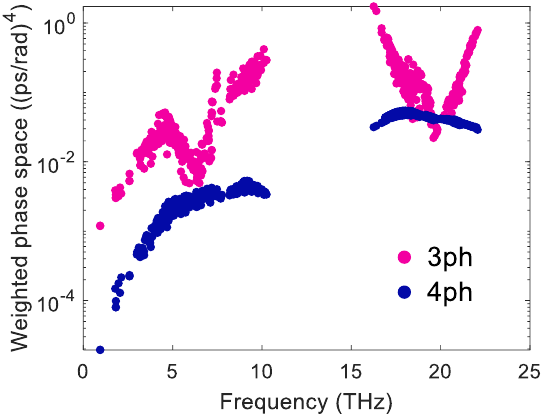}
    \caption{The weighted phase space of three-phonon and four-phonon scattering of GaN at room temperature.}
    \label{fig: Figure S1}
\end{figure}

As shown in Fig. \ref{fig: Figure S1}, the weighted phase space of four-phonon scattering is significantly smaller than that for three-phonon scattering. Therefore, the four-phonon scattering has negligible effects on the lattice thermal conductivity of GaN at room temperature.

\section{The Convergence tests on k-point mesh}
\begin{figure}[H]
    \centering
    \includegraphics[width=0.95\columnwidth]{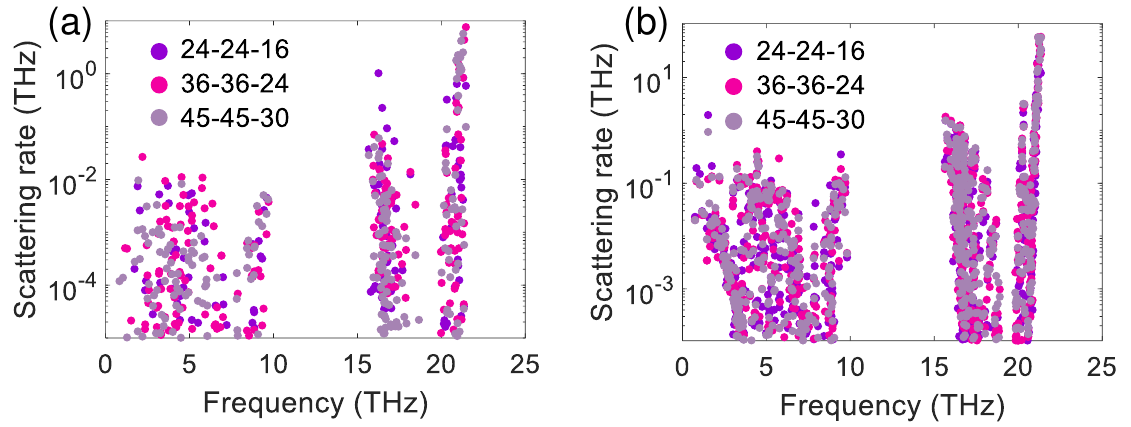}
    \caption{Phonon-electron scattering rates of (a) \textit{n}-type and (b) \textit{p}-type versus the different \textbf{k}-point meshes. The carrier concentrations are $10^{21} \, \text{cm}^{-3}$ in all the cases.}
    \label{fig: Figure S2}
\end{figure}

Since the phonon energy scale is much smaller than that of the electrons, A very dense electronic \textbf{k}-point mesh may be necessary to achieve convergence for the phonon-electron scattering rates. In Fig. \ref{fig: Figure S2}, we show the phonon-electron scattering rates of \textit{n}-type and \textit{p}-type GaN, with the \textbf{k}-point meshes of $24 \times 24 \times 16$, $36 \times 36 \times 24$ and $45 \times 45 \times 30$. It can be seen that phonon-electron scattering rates do not have significant variations using these \textbf{k}-point meshes. Therefore, we adopt $36 \times 36 \times 24$ \textbf{k}-point mesh in our calculations for GaN.

\section{Band structure of GaN without and with spin-orbit coupling (SOC)}
\begin{figure}[H]
    \centering
    \includegraphics[width=0.75\columnwidth]{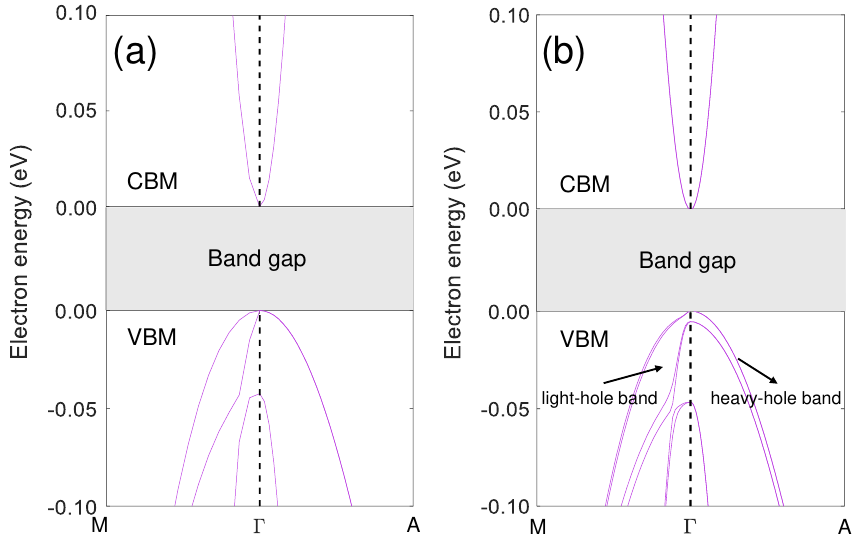}
    \caption{Band structure of GaN (a) without and (b) with SOC. The electron energy is aligned to the band edges.}
    \label{fig: Figure S3}
\end{figure}

Fig. \ref{fig: Figure S3} shows that the SOC has almost no effect on the conduction bands, while the light-hole and heavy-hole bands are split along the $\Gamma$-M high-symmetry path of the first Brillouin zone, resulting in a small split-off energy gap of 7 meV, agreeing well with the experimental value of 11 meV\cite{dingle1971absorption}.

\section{Phonon Dispersion of GaN}
\begin{figure}[H]
    \centering
    \includegraphics[width=0.52\columnwidth]{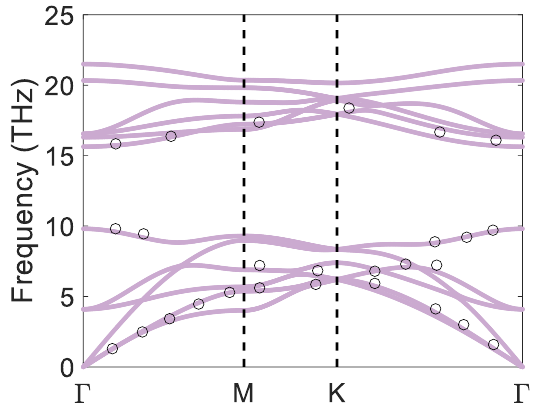}
    \caption{Phonon dispersion of GaN. The experimental data (symbols) are taken from Ref. \cite{ruf2001phonon}\@.}
    \label{fig: Figure S4}
\end{figure}

The phonon dispersion of GaN is shown in Fig. \ref{fig: Figure S4}. There are four atoms in the primitive cell of GaN. Correspondingly, there are twelve phonon branches. The phonon frequencies agree well with experimental data\cite{ruf2001phonon}.

\section{Phonon-phonon scattering rates of GaN and Si}
\begin{figure}[H]
    \centering
    \includegraphics[width=0.58\columnwidth]{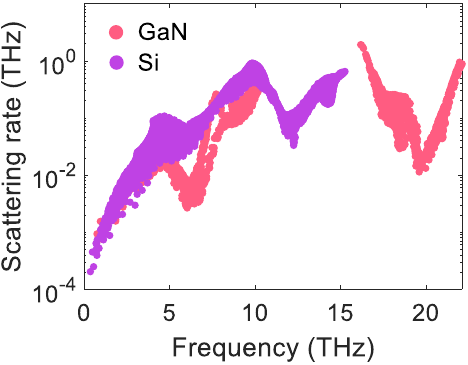}
    \caption{Phonon-phonon scattering rates of GaN and Si.}
    \label{fig: Figure S5}
\end{figure}

The phonon-phonon scattering rates of GaN and Si are shown in Fig. \ref{fig: Figure S5}. The phonon-phonon scattering rates of GaN for phonon frequencies below 10 THz are smaller than those of Si, resulting in the relatively larger lattice thermal conductivity of intrinsic GaN.

\section{The effects of carrier concentrations on phonon-electron scattering rates}
\begin{figure}[H]
    \centering
    \includegraphics[width=0.95\columnwidth]{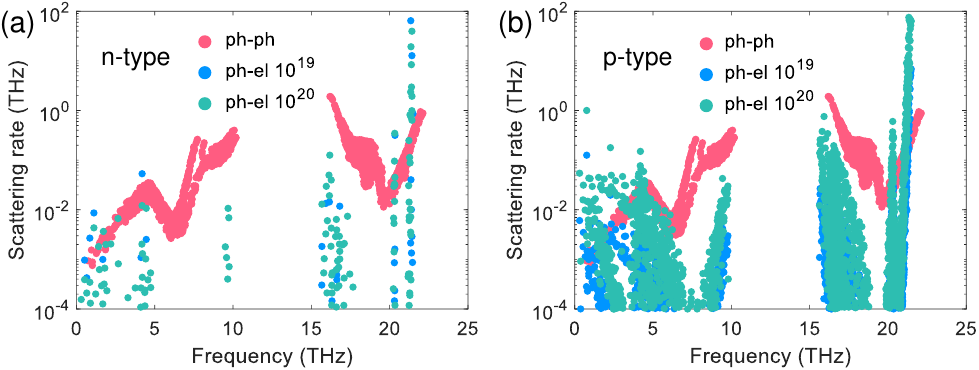}
    \caption{Phonon-electron scattering rates in (a) \textit{n}-type and (b) \textit{p}-type GaN at $10^{19}$ and $10^{20} \, \text{cm}^{-3}$ carrier concentrations with the phonon-phonon scattering rates.}
    \label{fig: Figure S6}
\end{figure}

Fig. \ref{fig: Figure S6} shows phonon-electron scattering rates at the carrier concentrations of $10^{19}$ and $10^{20} \, \text{cm}^{-3}$ for \textit{n}-type and \textit{p}-type GaN.

\section{Total and partial electron density of states from GaN}
\begin{figure}[H]
    \centering
    \includegraphics[width=0.52\columnwidth]{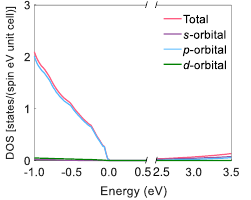}
    \caption{Total and partial electron density of states from GaN. The electron energy is normalized to the valence band maximum.}
    \label{fig: Figure S7}
\end{figure}

As shown in Fig. \ref{fig: Figure S7}, the electron DOS in the vicinity of the valence band maximum is primarily contributed by \textit{p} orbitals, which exhibit a triple degeneracy and minimal band dispersion. Conversely, the vicinity of the conduction band minimum is primarily contributed by \textit{s} orbitals, resulting in non-degenerate and dispersive electron bands.

\section{Fermi surface nesting functions for GaN and Si}
\begin{figure}[H]
    \centering
    \includegraphics[width=0.95\columnwidth]{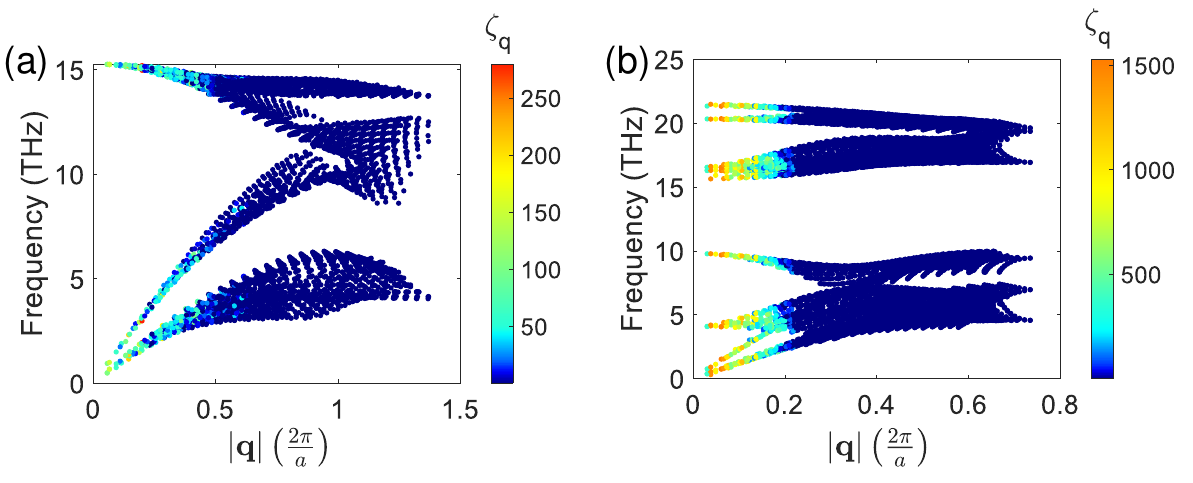}
    \caption{The Fermi surface nesting function for Si (a) and GaN (b) at room temperature. The carrier concentrations are $10^{21} \, \text{cm}^{-3}$ in all the cases.}
    \label{fig: Figure S8}
\end{figure}

As shown in Fig. \ref{fig: Figure S8}, the Fermi surface nesting function of GaN is larger than that of Si. Therefore, the relatively small reduction in $\kappa_{\text {lat}}$ of \textit{p}-type GaN cannot be explained by the Fermi surface nesting function and should be ascribed to the electron-phonon matrix elements.

\bibliography{bibliography}